%% file: main.tex
\documentclass[10pt,letterpaper]{article}
\usepackage[top=0.85in,left=1.2in,right=1.2in,footskip=0.75in,marginparwidth=2in]{geometry}
\usepackage{blindtext}
\usepackage{graphicx}
\usepackage[english]{babel}
\usepackage[utf8x]{inputenc}
\usepackage{mathtools}
\usepackage{amsmath}
\usepackage{url}
\usepackage{cuted}
\usepackage{arydshln}
\usepackage[numbers, sort]{natbib}

\usepackage[us]{datetime}
\usepackage{subfig}
\usepackage[normalem]{ulem}
\usepackage[ruled]{algorithm}
\usepackage{algpseudocode}

\makeatletter
\algnewcommand{\LineComment}[1]{\Statex \hskip\ALG@thistlm\textbf{arguments} #1}
\makeatother

\usepackage{soul} 
\usepackage{booktabs}

\PassOptionsToPackage{square,numbers}{natbib}
\usepackage[dvipsnames]{xcolor}

\newif\ifdraft
\drafttrue
\ifdraft
    \usepackage[normalem]{ulem}
    \usepackage[dvipsnames]{xcolor}
    \newcommand\remove[1]{\textcolor{red}{}}
    \newcommand\todo[1]{\textcolor{red}{~#1}}
    \newcommand\bertus[1]{\textcolor{orange}{~\textbf{Bertus:} #1}}
    
    \newcommand\maria[1]{\textcolor{purple}{~\textbf{Maria:} #1}}
    \newcommand\ando[1]{\textcolor{ForestGreen}{~\textbf{Ando:} #1}}
    \newcommand\marco[1]{\textcolor{ForestGreen}{~\textbf{Marco:} #1}}

\else
    \newcommand\remove[1]{}
    \newcommand\todo[1]{}
    \newcommand\bertus[1]{}
    
    \newcommand\maria[1]{}
    \newcommand\ando[1]{}
    \newcommand\marco[1]{}
\fi

\newcommand{\figref}[1]{Fig.~\ref{#1}}
\newcommand{\tabref}[1]{Table~\ref{#1}}
\newcommand{\eqnref}[1]{Eq.~\ref{#1}}
\newcommand{\secref}[1]{Section~\ref{#1}}
\newcommand{\appref}[1]{Appendix~\ref{#1}}
\newcommand{\algoref}[1]{Algorithm~\ref{#1}}

\newcommand{\rom}[1]{(\romannumeral #1)}
\newcommand{\q}[1]{\lq{#1}\rq}
\providecommand{\keywords}[1]{\bigskip \noindent\textbf{\textit{Index terms---}} #1}

\begin{document}

\begin{flushleft}
{\Large
\textbf\newline{Personalized advice for enhancing well-being using automated impulse response analysis --- AIRA}
}
\newline
\\
F. J. Blaauw\textsuperscript{a,b,c,\textdagger},
L. van der Krieke\textsuperscript{a},
A. C. Emerencia\textsuperscript{a,c},
M. Aiello\textsuperscript{b},
and P.~de~Jonge\textsuperscript{a, c}
\\
\bigskip
\textsuperscript{\textdagger} Corresponding author (f.j.blaauw@rug.nl)\\
\textsuperscript{a} University of Groningen, University Medical Center Groningen, University Center for Psychiatry, Interdisciplinary Center Psychopathology and Emotion regulation (ICPE), The Netherlands\\
\textsuperscript{b} University of Groningen, Johann Bernoulli Institute for Mathematics and Computer Science (JBI), Distributed Systems Group, The Netherlands\\
\textsuperscript{c} University of Groningen, Department of Developmental Psychology, The Netherlands
\bigskip

\end{flushleft}

\section*{Abstract}
The attention for personalized mental health care is thriving. Research data specific to the individual, such as time series sensor data or data from intensive longitudinal studies, is relevant from a research perspective, as analyses on these data can reveal the heterogeneity among the participants and provide more precise and individualized results than with group-based methods. However, using this data for self-management and to help the individual to improve his or her mental health has proven to be challenging.

The present work describes a novel approach to automatically generate personalized advice for the improvement of the well-being of individuals by using time series data from intensive longitudinal studies: \emph{Automated Impulse Response Analysis} (\textsc{aira}). \textsc{Aira} analyzes vector autoregression models of well-being by generating impulse response functions. These impulse response functions are used in simulations to determine which variables in the model have the largest influence on the other variables and thus on the well-being of the participant. The effects found can be used to support self-management.

We demonstrate the practical usefulness of \textsc{aira} by performing analysis on longitudinal self-reported data about psychological variables. To evaluate its effectiveness and efficacy, we ran its algorithms on two data sets ($N=4$ and $N=5$), and discuss the results. Furthermore, we compare \textsc{aira}'s output to the results of a previously published study and show that the results are comparable. By automating Impulse Response Function Analysis, \textsc{aira} fulfills the need for accurate individualized models of health outcomes at a low resource cost with the potential for upscaling.

\keywords{impulse response; automated; vector autoregression; simulation; personalized advice; electronic diary study data}

\section{Introduction} 
\label{sec:introduction}
Perspectives on \emph{eHealth} (computer aided health care) and \emph{mHealth} (mobile health) have changed greatly in the last decade~\cite{Meier2013, Fiordelli2013}. The modern individual possesses more mobile technology and \q{smart} devices than ever before~\cite{SpencerTraskandCo2014}. Some of these devices, such as smartphones and tablets, enable people to have Internet access during a large part of their daily life, allowing the use of new and more accurate methods to perform assessments in health care and health research~\cite{Trull2009}. Research for which a large number of assessments need to be conducted (possibly multiple times per day) can now be performed digitally using mobile technology. Technology enables researchers to carry out studies on a larger scale than would have been possible using traditional methods (i.e., collecting data using pencil and paper, using conventional postal methods for sending the data, and processing this data manually). Moreover, the use of technology can provide means for interactive and automated analysis methods. Using an automated method for analyzing the data might even be inevitable in large-scale studies. As vast amounts of data are collected for ever larger groups of people, the data might grow too large for manual analysis. Additionally, manual analysis can result in inconsistent or opinionated outcomes, which can be reduced by automatizing the procedure.

\subsection{Personalizing mental health research} 
\label{sub:personalizing_psychopathology_research}
In recent years, it has increasingly been acknowledged that mental health research requires a person-centered approach and should not focus only on group-based averages~\cite{RefWorks:85, Hamaker2012, Molenaar2013}. Since group-based analysis applies to the average person, several researchers have argued that conclusions from group-based analysis might not hold for \emph{any} individual~\cite{Lucas2006,RefWorks:84}. Instead of using a few data points of many people as the starting point for research, the focus needs to shift towards using many data points of individuals. This allows for studying changes within the individual (\emph{within-person variability}). Several studies have shown that mental health and ill-health are dynamic phenomena that vary highly between individuals and over time~\cite{RefWorks:4, Bouchard2007, VanderKrieke2015, VanderKrieke2015diary}. A technique to assess this within-person variability of mental health is the \emph{Ecological Momentary Assessment} method (\textsc{ema})~\cite{shiffman1998ecological}, in which a longitudinal data set is created by asking the individual to repeatedly complete the same assessment, for a certain period of time (e.g., in the morning, afternoon, and evening, for thirty days in a row).


\subsection{Analyzing diary study data} 
\label{sub:analyzing_time_series_data}
When questionnaires are filled out in sequence the data is a time series. A popular technique for analyzing multivariate, equally spaced time series data is \emph{vector autoregression} (\textsc{var})~\cite{RefWorks:5}. \textsc{Var} can be used to fit a multivariate regression model; a model in which the outcome of one variable (e.g., \q{concentration}) is regressed on the outcomes of several other variables (e.g., \q{self-esteem} and \q{agitation}). The \textsc{var} model itself is a set of such multivariate regression equations for a system of two or more variables, where each variable in the system is regressed on its own time-lagged values and the time-lagged values of all other variables in the system~\cite{Brandt2007}. That is, a variable $x$ at time $t$ is predicted by the same variable $x$ at time $t-1, t-2, \ldots, t-p$ and by other variables at time $t-1, t-2, \ldots, t-p$. The number of measurements used to look back in time ($p$) are called \emph{lags} in time series parlance.

\begin{figure}[ht]
  \centering
  \includegraphics[width=.9\columnwidth]{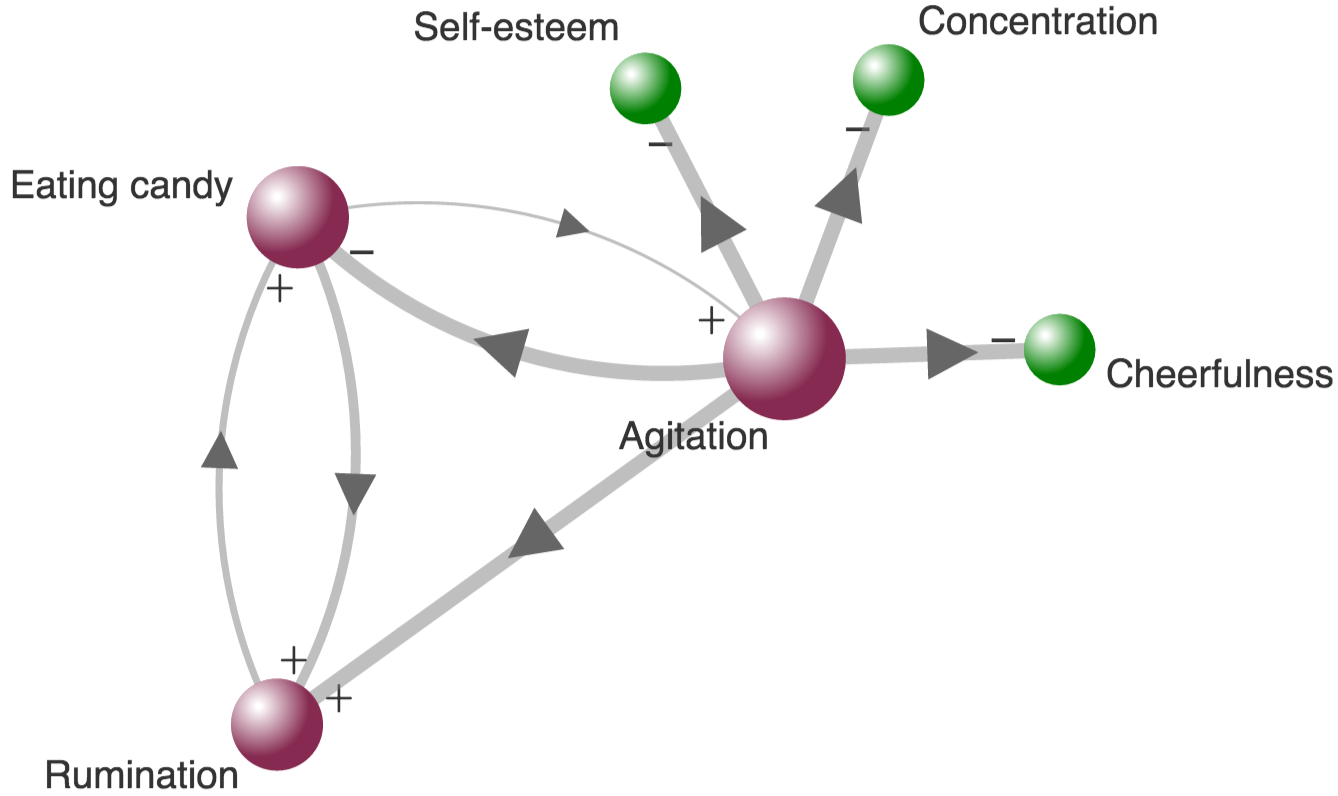}
  \caption{An example of a Granger causality graph, showing the relations between variables over time.}
\label{fig:images_dynamic}
\end{figure}

\textsc{Var} models allow for determining \emph{Granger causality}~\cite{Granger1969}. A variable $x$ \emph{Granger causes} another variable $y$, if and only if the variance of $y$ can be better explained by lagged values of $y$ and lagged values of another variable ($x$), than lagged values of $y$ alone~\cite{Granger1969}. Granger causal relations can be depicted by means of a weighted directed graph. \figref{fig:images_dynamic} gives an example of such a graph, in which the nodes represent the variables (which can be self-reported psychological variables, such as measured in the present study, physiological variables as read from sensors, or other variables), the connections (or edges) represent the significant directed relations between the variables and the thickness of an edge represents the strength of the relation. In this image, the green nodes represent variables that participants usually experience as a positive phenomenon and the red nodes represent variables that participants usually experience as a negative phenomenon. In \figref{fig:images_dynamic}, for example, an increase in the variable \q{agitation} at time $t-1$ Granger causes an increase in \q{rumination} and a decrease in \q{self-esteem}, \q{concentration}, \q{cheerfulness}, and \q{eating candy} at time $t$. The figure only shows relations present at lag 1.

Graphs like the one depicted in \figref{fig:images_dynamic} encompass relevant information regarding the interactions in a \textsc{var} model that could be of interest to the participant. However, these graphs lack several important features to serve as a means to provide advice on how to improve the participant's well-being. Firstly, participants may have a hard time understanding these graphs~\cite{VanderKrieke2015diary}. This can be attributed to the conceptual complexity of the different edge and node types in the graph. Secondly, these graphs give a general overview of the coefficients in a \textsc{var} model by providing an edge-focused representation. Although such a representation gives information about the individual relations between nodes, it remains complicated to interpret the model as a whole, especially with respect to the temporal interplay between the nodes in the model. The \textsc{var} coefficients are meaningless to interpret individually, as it is the \textsc{var} model as a whole that describes the complete dynamic behavior of the variables in the system~\cite{Brandt2007}. 

\subsection{Automated Impulse Response Analysis}
\label{sub:advanced_impulse_response_analysis}
In the present work, we describe \emph{Automated Impulse Response Analysis} (\textsc{aira}), an approach to automatically generate advice for improving a participant's well-being using \textsc{var} models derived from \textsc{ema} data. \textsc{Aira} creates advice by simulating the interactions between variables in a \textsc{var} model (i.e., showing what would happen to $y$ when variable $x$ increases). The technique \textsc{aira} uses is called \emph{Impulse Response Function} (\textsc{irf}) analysis. \textsc{Irf} analysis allows us to \emph{shock} (that is, give an instantaneous exogenous impulse to) certain variables to see how this shock propagates through the various (time-lagged) relations in the \textsc{var} model. In other words, \textsc{irf} shows how variables respond to an impulse applied to other variables~\cite{Brandt2007}. \textsc{Aira} generates the impulse response functions for each of the equations in a \textsc{var} model, and analyzes these \textsc{irf}s to automatically generate personalized advice. \textsc{Aira} uses and partly extends some of our previous work; the automatic creation of \textsc{var} models~\cite{Emerencia2013}. The fact that \textsc{aira} analyzes the \textsc{var} model as a whole enables \textsc{aira} to be a more appropriate and more precise technique for analyzing a \textsc{var} model than a mere manual inspection of said model. A second novelty of the present work is an implementation of \textsc{var} and \textsc{irf} analysis in the JavaScript-language. As far as we know, this is the first openly available cross-platform web-based implementation of its kind. The JavaScript implementation can be useful for calculating \textsc{var} models or \textsc{irf}s in the browser, or on a server running for example \emph{NodeJS\footnote{Website: http://nodejs.org}}, which can aid upscaling.

\textsc{Aira} generates several types of advice answering the following questions: \rom{1} \emph{Which of the variables has the largest effect on my well-being?}, \rom{2} \emph{How long is $Y$ affected by an increase in $X$?}, and \rom{3} \emph{What can I do to change a certain $Y$ variable?} Firstly, \textsc{aira} shows how well each of the modeled variables can be used to improve all other variables in the network, by summing over the effects variables have on the other variables. For this type of advice, we consider an improvement of the complete network an improvement of the participant's well-being in general. Secondly, \textsc{aira} can provide insight into the duration of an effect, giving insight in the persistence of a perceived relation. Thirdly, \textsc{aira} allows the participant to select a variable he or she would like to improve and by how much, after which \textsc{aira} will try to find a suitable solution to achieve this improvement. \textsc{Aira} will iterate over all other variables and estimate for each of these variables how large a change is needed to achieve the desired effect.

This paper is organized as follows. \secref{sec:related_work} gives an overview of related work. \secref{sec:variable_selection} illustrates the concept of \textsc{aira} presenting its mathematical foundation. \secref{sec:implementation} describes \textsc{aira} by presenting pseudocode of the algorithms. \secref{sec:experimental_results} describes the experimental results acquired when evaluating \textsc{aira}. We also evaluate the implementation of \textsc{aira} by comparing its analysis with a manually performed analysis. \secref{sec:discussion} discusses the results. \secref{sec:conclusion_and_future_work} concludes the work and provides direction for future work.


\section{Related work} 
\label{sec:related_work}
\textsc{Irf} is a technique used in several fields, including digital and analog signal processing~\cite{Bellanger2000}, control theory~\cite{Hespanha2009}, psychiatry~\cite{Hoenders2011}, econometrics~\cite{Pesaran1997}, and even for the quality assurance of fruit~\cite{Diezma-Iglesias2004}. Each field applies \textsc{irf} in a differently, but they all revolve around a comparable principle. \textsc{Irf}s are used to determine how a model or system, or part of that model or system, responds to a sudden large change or shock on one or more of the variables.

\subsection{Personalized feedback on EMA data}
\label{sub:personalized_feedback_on_ema_data}
Several applications exist that can provide users with feedback based on diary studies or other longitudinal health data collected by them. The feedback and advice provided in \textsc{ema} studies can roughly be split into two categories: \emph{real-time advice} and \emph{post-hoc advice}. The type of advice chosen for a study depends on the goal of the study.

Real-time advice can be used to intervene directly in the \textsc{ema} study. For example, \citeauthor{Hareva2009} present advice to a participant by means of applying a severity threshold to the \textsc{ema} results~\cite{Hareva2009}. They describe a use case of their system for smoking cessation in which an email is automatically sent whenever a participant has smoked fewer cigarettes than a set threshold in order to encourage the behavior of smoking less. Real-time advice triggered by diary data has also been applied in treating childhood overweight~\cite{Bauer2010}. In a study by~\citeauthor{Bauer2010}, juvenile participants sent weekly text messages (\textsc{sms}) describing various parameters (such as emotion and eating behavior), after which an algorithm would automatically compare these results to the preceding week and create advice~\cite{Bauer2010}. Furthermore, \citeauthor{Ebner-Priemer2009} reviewed the use of diary studies in psychopathology research~\cite{Ebner-Priemer2009}. They discuss \textsc{ema} studies about mood-disorder and mood dysregulation (mainly bipolar disorder, depression, and borderline personality), and also methods for providing feedback on these data. 

Post-hoc advice is advice offered after completing the study. One of the advantages of this method is that elaborate statistical analysis can be performed, since all collected data can be used for analysis (instead of a small window of data). Furthermore, post-hoc feedback can be used in cases where the goal is to map the baseline behavior of a participant, not affected by interventions such as real-time feedback. To the best of our knowledge, only a few studies exist to date that provide personalized, post-hoc advice based on the dynamic relations between the variables in an \textsc{ema} study. In an electronic diary study performed by \citeauthor{Booij2015} participants received a post-hoc personal report on their daily activity and mood patterns~\cite{Booij2015}. \citeauthor{VanRoekel2016} provided participants of their electronic diary study with a written report and gave them lifestyle advice based on \textsc{var} models and the variables most promising for inducing a change in pleasure~\cite{VanRoekel2016}. Two of our own studies that provide such post-hoc advice are \emph{HowNutsAreTheDutch}~\cite{Blaauw2014, VanderKrieke2015} and \emph{Leefplezier}~\cite{Blaauw_leefplezier_2014}. HowNutsAreTheDutch is a Dutch mental health study that, inter alia, offers an \textsc{ema} study. In HowNutsAreTheDutch, a participant receives post-hoc feedback mainly consisting of a graph as shown in \figref{fig:images_dynamic} combined with a static description text. The second example, \emph{Leefplezier} (English: `Joys of living'), is a study focusing on the elderly population of the Netherlands. During the study, participants receive minimal feedback consisting of basic charts intended to motivate (but not influence) the users to complete the next assessments. At the end of a study, Leefplezier participants who finished with enough completed measurements, receive feedback comparable to that received by the HowNutsAreTheDutch participants. Both studies provide feedback based on \textsc{var} models, but merely present Granger causality networks as-is (as shown in \figref{fig:images_dynamic}), without performing analysis on the \textsc{var} models. Furthermore, the feedback provided in these studies is mainly descriptive and does not provide concrete examples on how participants can enhance their well-being, unlike \textsc{aira}.

\subsection{Vector autoregression in psychopathology research} 
\label{sub:vector_autoregression_in_psychopathology_research}
Several studies exist in mental health research in which data is longitudinally collected from individuals. Often, the analysis in these studies involves techniques designed for group studies, instead of purely individual-based analysis methods~\cite{Molenaar2013}. These methods are fine to use for drawing conclusions at the general level, but they might not hold at the individual level~\cite{Lucas2006,RefWorks:84}.

One analysis method applicable for \textsc{ema} studies focusing purely on the individual is \textsc{var}. The use of \textsc{var} in mental health research is relatively new. Several researchers have incorporated this type of analysis (e.g., \citeauthor{Dugas2009}~\cite{Dugas2009}, \citeauthor{Snippe2014}~\cite{Snippe2014} and \citeauthor{VanGils2014}~\cite{VanGils2014}), including some of the authors of the present work~\cite{Hoenders2011,RefWorks:4}).

The manual creation of \textsc{var} models can be challenging; for example, researchers have to select the appropriate variables and decide upon lag length. In ~\cite{Emerencia2013}, we describe a method to automate \textsc{var} modeling named \emph{Autovar}, which can solve these issues by selecting some of the optimal parameters for the model configuration. Autovar can automatically create and evaluate \textsc{var} models based on \textsc{ema} data. The models offered by Autovar adhere to assumptions of a valid \textsc{var} model, such as the stability assumption, the white noise assumption, the homoskedasticity assumption, and the normality assumption~\cite{Emerencia2013}. It reveals the Granger causality found in the data and generates a network graph. However, Autovar does not generate specific advice nor does it apply \textsc{irf} to the created models.The creation of \textsc{var} models themselves, is outside the scope of the present work, see for instance~\cite{Brandt2007} and~\cite{Emerencia2013}.


\section{From Variable Selection to Advice Generation} 
\label{sec:variable_selection} 
\begin{figure}[htbp]
  \centering
  \includegraphics[width=\columnwidth]{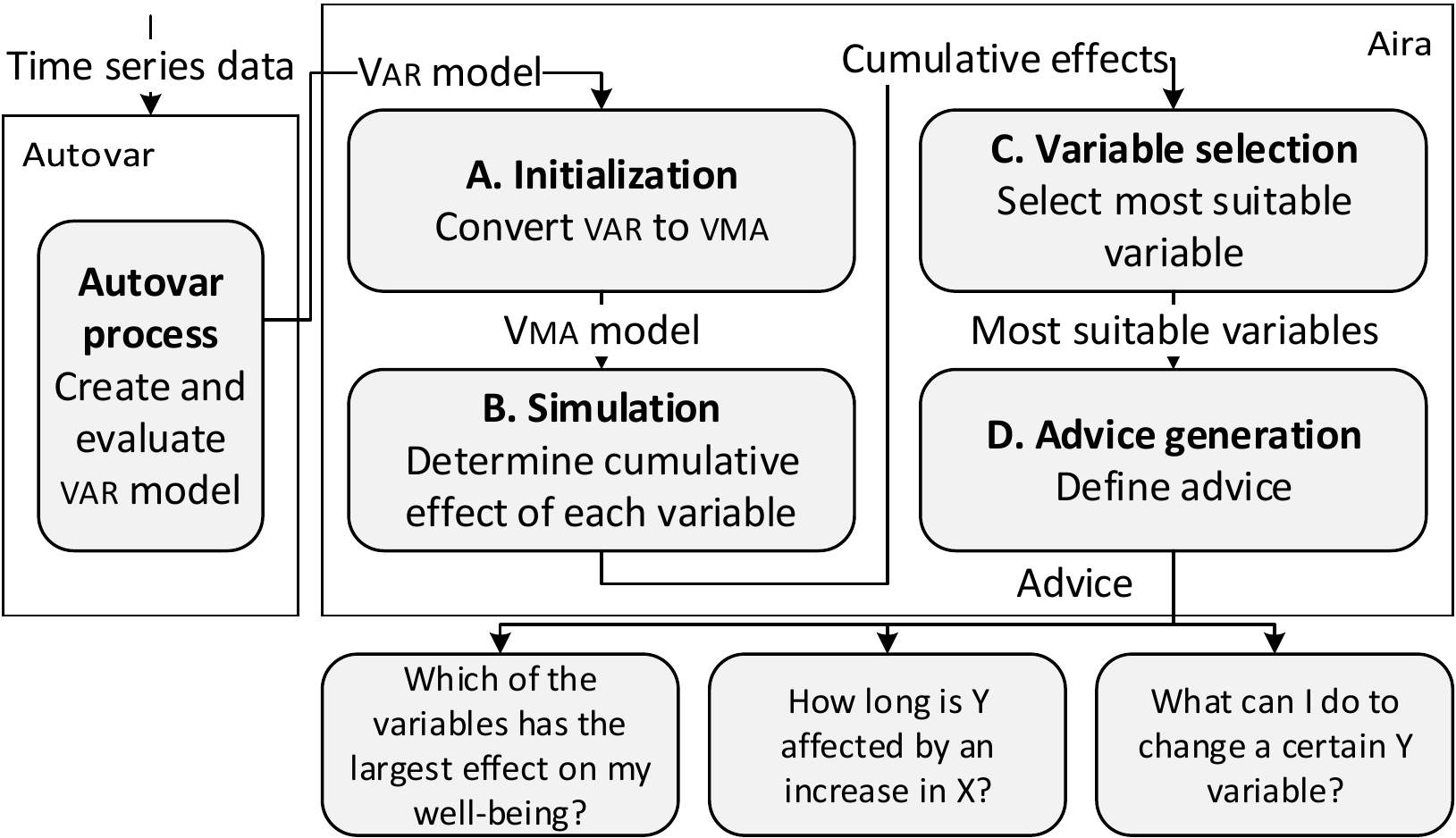}
  \caption{Overview of of the AIRA advice generation process.}
\label{fig:aira-process}
\end{figure}

\textsc{Aira} uses \textsc{irf} analysis to determine the effect each variable has on the other variables in the model for generating advice. The outcome of this analysis is then converted by \textsc{aira} to several types of advice for the participant. The advice describes in several ways which of the variables can best be adjusted in order to achieve the desired effect. The advice generation process of \textsc{aira} can be subdivided into four phases: (a) initialization, (b) simulation, (c) variable selection, and (d) advice generation. The \textsc{aira} process is illustrated in \figref{fig:aira-process}.

\subsection{Initialization} 
\label{sub:initialization}
In the first phase, \textsc{aira} converts the \textsc{var} model into a \emph{vector moving average} (\textsc{vma}) representation. This \textsc{vma} representation shows how the model responds to changes in the residuals or to exogenous shocks on the model (i.e., shocks from outside of the model)~\cite[p.~37]{Brandt2007}. As an example, \eqnref{eq:var-representation} shows a basic $var(p)$ model (a \textsc{var} model with $p$ lags), \eqnref{eq:vma-representation} shows the $vma(\infty)$ representation of the same model.

\begin{subequations}
  \begin{align}
    Y_{t} &= c + B^{1}Y_{t-1} + \cdots + B^{p}Y_{t-p} + \xi X+ \vec{e}_t
    \label{eq:var-representation}\\
    Y_{t} - d &= \vec{e}_t \cdot (I_m + (\sum_{i=1}^k C_{1,i}) L + (\sum_{i=1}^k C_{2,i}) L^{2} + \cdots)
    \label{eq:vma-representation}
  \end{align}
  \label{eq:var-and-vma-representation}
\end{subequations}

\begin{figure*}[ht!]
  \normalsize

  \begin{equation}
    C =
    \begin{bmatrix}
      B^1          & 0          & 0          & 0        & 0         & 0                     & 0                                         & 0         & \cdots\\
      B^1C_1       & B^2        & 0          & 0        & 0         & 0                     & 0                                         & 0         & \cdots\\
      B^1C_2       & B^2C_1     & B^3        & 0        & 0         & 0                     & 0                                         & 0         & \cdots\\
      B^1C_3       & B^2C_2     & B^3C_1     & B^4      & 0         & 0                     & 0                                         & 0         & \cdots\\
      \vdots & \ddots &  \\
      B^1C_{p-1} & B^2C_{p-2}  & B^3C_{p-3} & \cdots & B^{p-1}C_{1} & B^{p}                 & 0                                         & 0         & \cdots\\
      \vdots & \ddots &  \\
      B^1C_{\infty-1} & B^2C_{\infty-2} & B^3C_{\infty-3}& \cdots & B^{p-1}C_{\infty-(p-1)} & B^pC_{\infty-p} & 0\cdot C_{\infty-(p+1)} & \cdots    & \cdots\\
    \end{bmatrix}
    \label{eq:c_matrix}
  \end{equation}
\end{figure*}

In \eqnref{eq:var-representation}, $Y_{t}$ is a vector containing $m$ variables ($\vec{v}$) at time $t$, $p$ is the number of lags in the model (i.e., the number of previous measurements included in the model for the prediction of the current variable), $c$ is a vector containing $m$ constant terms, $X$ are the exogenous variables (i.e., variables that can influence the model, but cannot be influenced by the model, such as the day of the week or the weather), $\xi$ is the coefficient matrix for the exogenous variables, $\vec{e}_{t}$ is a vector containing the error in the model (i.e., all variance left in the data not explained by the model), and each $B^1$, $B^2$, $\cdots$, $B^p$ is a coefficient matrix for a specific lag in time. Each entry ($\beta$) of one of the matrices $B^p$ is a coefficient for one variable predicting another value at a specific lag. Each matrix in $B^p$ has the order variable $\times$ coefficients. That is, each \emph{row} of a $B^p$ matrix represents the coefficients of lag $p$ for the variable in that row (e.g., $\beta_{i,j}^p$ is the scalar coefficient for predicting variable $i$ using variable $j$, at lag $p$).

\eqnref{eq:vma-representation} shows the same model as \eqnref{eq:var-representation}. However, in this case the model is recentered around its equilibrium values, and it is converted to a function of the error term in the model~\cite{Brandt2007}. The \textsc{vma} representation allows one to investigate the standardized impact of a shock on the model. For more information on the \textsc{var} to \textsc{vma} model conversion, see  \citeauthor{Brandt2007}~\cite{Brandt2007}. In \eqnref{eq:vma-representation}, $L^k$ is the lag operator which shifts the variable that it is multiplied by with $k$ steps, i.e., $L^k x_t = x_{t-k}$. $C_1$, $C_2$, etc., are the \textsc{vma} coefficient matrices of the model. Each $C_i$ represents the $i^{th}$ row of $C$, each $C_{i,j}$ a specific element from that row. $C$ itself is a block lower triangular matrix, as shown in \eqnref{eq:c_matrix}. $C$ is contained in itself, as each row of $C_i$ contains all preceding entries in $C$ (i.e., $C_{i-1}$, $C_{i-2}$ until $C_0$). The reason for this recursion is that an effect at time $t$ is also affected by all of the preceding effects ($t-1$, $t-2$, etc.). Theoretically, the $C$ matrix can contain an infinite number of rows. However, when using a stable \textsc{var} model the responses will eventually converge to zero~\cite{Brandt2007}. This number of rows can be limited to a certain number of steps ($k$), also known as \emph{horizon}. The horizon is the total number of steps the \textsc{irf} model will predict. The $\vec{e}_t$ vector ($m\times 1$) originates from an $m \times k$ matrix $E$ that contains the error terms for each variable corresponding to a variable at the same location in $\vec{v}$. When performing \textsc{irf} analysis, the $\vec{e}_t$ term is replaced by a vector of shocks $\vec{s}$, which is a not-lagged vector of structural shocks to the model. In this work, we assume shocks of unit size, i.e., each entry in this vector is either $1$ (if a variable receives a shock) or $0$ (if a variable does not receive a shock). Since the effects are standardized, this corresponds to a $\sigma$ (standard deviation) increase.


\eqnref{eq:vma-representation} does not take into account the contemporaneous effects, and captures these effects in the error term. It is often problematic to determine the direction of the effect for the contemporaneous effects, as they merely show the correlation, and therefore the direction remains unclear. If one has a hypothesis regarding the directionality of these contemporaneous effects they could be implemented  by using a technique named \emph{orthogonalized \textsc{irf}} (\textsc{oirf})~\cite{Brandt2007}. In that case, $I_m$ (the $m \times m$ identity matrix) should be changed to a matrix representing the contemporaneous effects, e.g., by computing the first matrix from the Cholesky decomposition of the error covariance matrix and possibly testing all directions for the effects, or by using theoretical domain knowledge for the directions~\cite{Pesaran1997, Brandt2007}. $d$ is the \textsc{var} constant term divided by the vector autoregressive lag polynomial.

\subsection{Simulation} 
\label{sub:simulation}
In the second phase, \textsc{aira} runs \textsc{irf} calculations for each of the variables in the model. \textsc{Aira} simulates an impulse on one variable ($x$) in the model and registers how all other variables respond. The response of each variable is used to determine the effect of a variable on the other variables in the model. Besides using all responses caused by an impulse, \textsc{Aira} can also be configured to consider only the effects which have a certainty of at least $95\%$, by bootstrapping the model~\cite{RefWorks:3}. The methods in this phase allow for determining the variables most suitable for influencing other variables and serves as a preprocessing step for generating advice.

\begin{figure}[htbp]
  \includegraphics[width=\columnwidth]{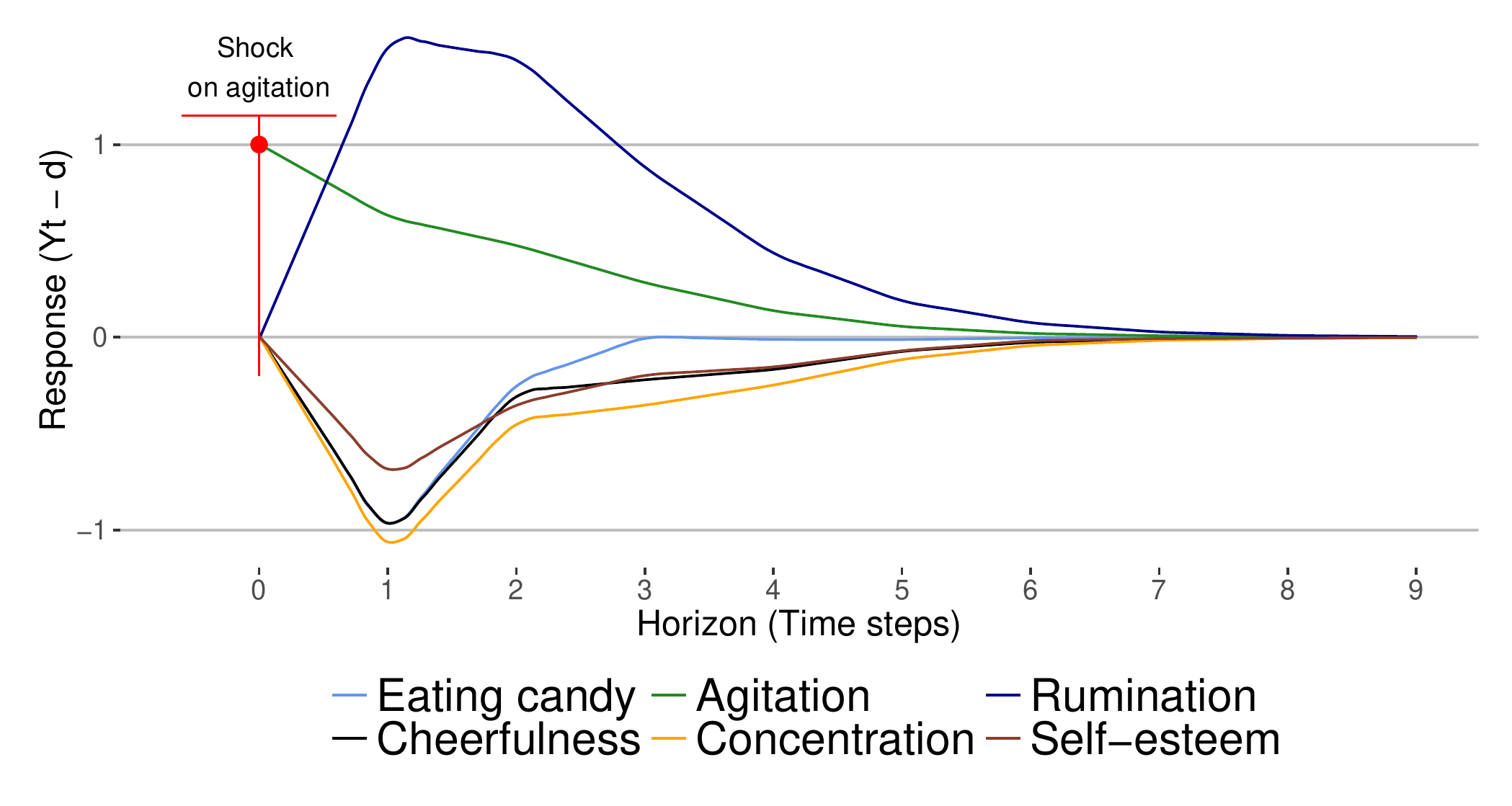}
  \caption{An example of the responses of six variables from a $var(1)$ model after a shock on the variable \q{agitation}, corresponding to the network shown in \figref{fig:images_dynamic}.}
\label{fig:images-example-response}
\end{figure}

An example of an \textsc{irf} of the network model used in \figref{fig:images_dynamic} is shown in \figref{fig:images-example-response}. The figure shows how a shock on \q{agitation} (the green line) affects the other variables in the model. The shock is a $\sigma$ increase of the variable \q{agitation} (at time $t=0$). The other variables respond from $t=1$ onwards (as contemporaneous relations are neglected in this example). The shock on \q{agitation} has an effect on most of the variables in the model: a negative effect on \q{eating candy}, \q{cheerfulness}, \q{self-esteem}, and \q{concentration}, and a positive effect on \q{rumination}. The effect vanishes after $\approx 7$ time steps.

The response of one variable to changes in another variable can be calculated using \eqnref{eq:irf-function}, in which $x$ is the index of the variable receiving the shock and $y$ the index of the variable for which the response is analyzed. The outcome of this equation is a vector with the response of $y$, where each entry is a response on the horizon ($k$). The remainder of the equation is similar to \eqnref{eq:vma-representation}, however, in this equation, $\vec{\alpha}$ is a vector of zeros, with a $1$ on the index of the variable to shock ($x$).

\begin{equation}
    \begin{aligned}
    irf(x, y, k) &= [(\psi_0 \cdot \vec{\alpha}(x))_y, \ldots, (\psi_k \cdot \vec{\alpha}(x))_y]^T\\
    \psi_i &= 
    \begin{cases}
        I_m   & \text{if } i = 0\\
        \sum_{j=1}^{i}C_{i,j} & \text{if } i \ge 1
    \end{cases}
  \end{aligned}
  \label{eq:irf-function}
\end{equation}

\begin{figure}[htbp]
  \includegraphics[width=.9\columnwidth]{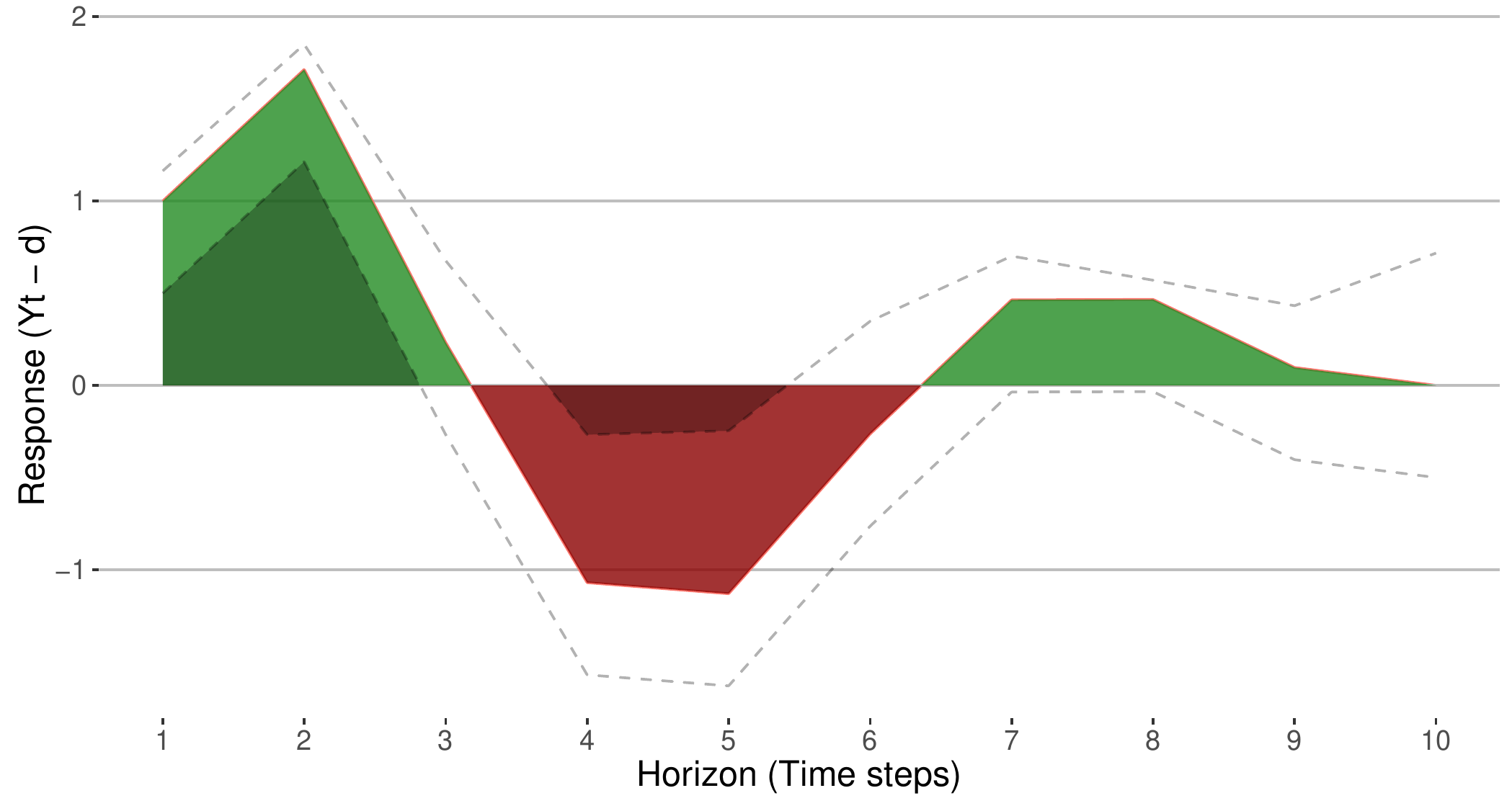}
  \caption{Artificial example of the \textsc{auc} to demonstrate the response to an impulse.}
\label{fig:pos-neg-area}
\end{figure}

In addition, \textsc{aira} applies a form of \emph{cumulative \textsc{irf}} analysis to determine the total response a variable has on another variable. In cumulative \textsc{irf} analysis, the response of a variable is summed to a total value, as shown in \eqnref{eq:cumulative-sum}. The cumulative \textsc{irf} is equivalent to the net area under the curve (\textsc{auc}), where areas corresponding to a response less than zero are subtracted from areas corresponding to responses higher than zero. Using the example \textsc{irf} in \figref{fig:pos-neg-area}, the cumulative response is the green areas minus the red areas.  The definition of the \textsc{irf} as shown in \eqnref{eq:irf-function} takes all responses into account, which might be too optimistic and cause inaccuracies. These inaccuracies may cause small insignificant effects to add up to a large, seemingly significant effect. To circumvent this, \textsc{Aira} can be configured so that it only considers significant effects by bootstrapping the results~\cite{Brandt2007,RefWorks:3,Sims1999} and only use effects significantly different from $0$. In \figref{fig:pos-neg-area} the darker areas depict the significant areas. The dashed lines indicate the 95\% confidence interval.

\eqnref{eq:cumulative-sum} shows the calculation of the cumulative \textsc{irf}, where $x$ is the variable that is shocked and $y$ is the variable the response is analyzed on. It sums all responses on the horizon $k$. The advantage of using this cumulative approach is that we obtain a single value representing the total effect of a single variable on another variable in the model, whilst taking into account all other interrelated variables.

\begin{equation}
  irf_{cum}(x, y, k) = \sum_{j=0}^k irf (x, y, k)_j
  \label{eq:cumulative-sum}
\end{equation}


\subsection{Variable selection} 
\label{sub:variable_selection}
In the third phase, \textsc{aira} selects the variable that is most suitable for adjusting the other variables in the model. \textsc{Aira} determines the net effect one variable has on all other variables. By using the total \textsc{auc}, including the negative effects, \textsc{aira} gives an estimate of the net effect a variable has. The function for calculating the net effect of a single variable ($x$) is shown in \eqnref{eq:total-irf}. In \eqnref{eq:total-irf} $k$ is the horizon over which the effect is calculated, and $m$ is the number of variables in the model. The result of this equation is the net effect variable $x$ has on the other variables. This result is contained in a vector of size $m-1$.

\begin{equation}
    irf_{total}(x, k) = \sum_{\substack{i=1\\i \neq x}}^m irf_{cum}(x, i, k)
    \label{eq:total-irf}
\end{equation}

In \eqnref{eq:total-irf}, the \textsc{irf}$_{cum}$ function uses a \textsc{var} model that handles \q{positive} and \q{negative} variables differently. Whether a variable is \q{positive} or \q{negative} is defined by its interpretation, that is, variables are considered positive or negative with respect to the well-being of a participant. For instance, a model might include two variables: \q{agitation} and \q{cheerfulness}, in which \q{Agitation} is considered a negative variable and \q{cheerfulness} is interpreted as a positive one. A variable deemed positive (\q{cheerfulness}) is presumably preferred to be increased, while a negative variable (\q{agitation}) is preferred to be decreased. To deal with this dichotomy, a transformation is performed on the negative variables using the $\Gamma$ and $E$ matrices; two matrices that convert the variables of a \textsc{var} model so that they are always positive. That is, negative variables change sign so their interpretation switches from an increase to a decrease of said variable. Each entry of the $\Gamma$ matrix is $\in \{1,-1\}$ and is created using $\vec{vlabels}$, a vector representing the interpretation of a variable. That is, $\vec{vlabels}_i = 1$ if the variable on position $i$ in $\vec{v}$ (a vector containing the variables in the model) is positive, and $\vec{vlabels}_i = -1$ if $\vec{v}_i$ is negative. The $E$ matrix is a matrix of which the rows of a negative endogenous variable are negative. \eqnref{eq:gamma-matrix} shows the creation of these matrices. This example shows $\vec{vlabels}$ built using a $\vec{v}$ with $6$ elements: $\vec{v} = $ [\q{eating candy (negative)}, \q{cheerfulness (positive)}, \q{agitation (negative)}, \q{concentration (positive)}, \q{rumination  (negative)}, \q{self-esteem (positive)}]$^T$. The $l$ variable denotes the number of exogenous variables in the model. After the transformation the $\vec{vlabels}$ can be considered a vector of all ones.

\begin{equation}
  \begin{aligned}
    \vec{vlabels} &= [-1,1,-1,1,-1,1]^T\\
    \Gamma &= \vec{vlabels} \cdot \vec{vlabels}^T\\
    E &= \vec{vlabels} \cdot \vec{1}_l^T
  \end{aligned}
  \label{eq:gamma-matrix}
\end{equation}

These $\Gamma$ and $E$ matrices are used to calculate the Hadamard product of $\Gamma$ and the $B$ matrices ($\Gamma \circ B^{p}$) and $E$ and the $\xi$ matrices ($E \circ \xi$) of \eqnref{eq:var-representation}. This equation is then used as input for the $irf_{cum}$ function. \textsc{Aira} can now determine the total effect of each variable on other variables, and therewith calculate the effect of a variable on the network as a whole.

\subsection{Advice generation} 
\label{sub:advice_generation}
In the last phase, \textsc{aira} generates the actual advice for the participant. The procedures of the previous phases are combined and personalized advice is constructed. \textsc{Aira} currently generates three types of advice: \rom{1} most influential node in network, \rom{2} percentage effect, and \rom{3} length of an effect.

\subsubsection{Most influential node in network}
\label{subsub:most_influential_node_in_network}
\textsc{Aira} identifies the variable with the largest net positive effect that can best be used for changing other variables. The advice describes how a participant can have the largest positive influence on his or her well-being. For example, if we determine that overall an increase in activity has a positive effect on the network, the advice would be: \q{If you were to increasing your amount of activity, this seems to positively affect your well-being}. The well-being of a participant is expressed by all variables in the network model. An increase in the network as a whole is considered an increase in the well-being of the participant. The generated advice consists of a sorted list of all variables and the extent to which they positively (i.e., increase one's well-being), negatively (i.e., decrease one's well-being), or neutrally (i.e., do not affect one's well-being) impact the network. The net effect of a variable is calculated by summing the cumulative \textsc{irf}s, as described in \eqnref{eq:total-irf}. By using the net effect approach, we offer a method to deal with the issue where different lags of a variable have conflicting signs for effects. For example, a variable can have a positive coefficient for a variable on the first lag, but a negative effect in the second one. \textsc{Aira} balances these effects by using the net effect.

\subsubsection{Length of effect}
\label{subsub:length_of_effect}
\textsc{Aira} shows the participant how long an impulse is estimated to have an effect on the other variables in the model. For calculating the length of an effect \textsc{aira} uses the (bootstrapped) \textsc{irf} as input, and determines how long, in minutes, a response of a variable remains (significantly) different from zero. The length is calculated by multiplying the \textsc{EMA} measurement interval with the number of steps on the horizon for which the effect is (significantly) greater or less than zero. For example, if an impulse on \q{activity} has an effect smaller than zero on \q{depression} for two time-steps and the measurement interval is $6$ hours, \textsc{aira} would state that a one standard deviation increase on \q{activity} has a negative effect on \q{depression} for approximately $720$ minutes (or $12$ hours). Furthermore, it determines how long the effective horizon is with respect to the given impulse. That is, after how many steps all the (significant) effects have converged to zero.

\subsubsection{Percentage effect}
\label{subsub:percentage_effect}
\textsc{Aira} generates specific advice showing what a participant could do in order to improve a single specific variable in the network by a specified percentage. For example, imagine a participant that would like to increase his or her \q{cheerfulness} variable by $10\%$. \textsc{Aira} can determine how to achieve this increase by advising the participant to either increase or decrease certain other variables in the model. Advice could then be generated as follows: \q{In order to increase your \emph{cheerfulness} by $10\%$, you can either increase your \emph{concentration} by $20\%$ or decrease your \emph{agitation} by $33\%$.} In order to calculate such advice, \textsc{aira} uses \eqnref{eq:percentage-effect}, provided that $irf_{cum}(y,x) \neq 0$, $\hat{y} \neq 0$, and $\sigma_{x} \neq 0$.

\begin{equation}
    \Delta_y = \frac{(\hat{k} \cdot \hat{x}) \cdot \Delta \cdot \sigma_y}{\hat{y} \cdot irf_{cum}{(y, x, k)} \cdot \sigma_x} \ (\forall y \mid \vec{v}_y \in \vec{v},\ y \neq x )
    \label{eq:percentage-effect}
\end{equation}

In \eqnref{eq:percentage-effect}, $\Delta \cdot 100$ is the desired percentage for increasing ($\Delta > 0$) or decreasing ($\Delta < 0$) variable $x$. $\Delta_y \cdot 100 $ is the calculated percentage effect $y$ has on $x$. $\hat{x}$ and $\hat{y}$ represent the average scores for the variables $x$ and $y$ of a participant respectively, $\sigma_x$ and $\sigma_y$ are the standard deviations of respectively variables $x$ and $y$, and $\hat{k}$ is the effective horizon over which the effect is calculated. The effective horizon is the number of steps on the horizon when the response of a shock has not yet converged to zero. The calculation is performed for all variables in the model ($\vec{v}_y \in \vec{v}$) not equal to the variable to improve ($y \ne x$), and the advice is established from the outcomes for each variable. The total effect $y$ has on $x$ is used in the calculation.

As an example, assume a person having a variable $x$ (the variable a person would like to change) with a mean value of $50$, and $\sigma_x = 15$. Assuming normality, an increase of one standard deviation would mean an increase of $15$, i.e., an increase of $\frac{\sigma_x}{\hat{x}} * 100 = 30\%$. Assume this person would like to increase the value of $x$ by $10\%$, we would require an increase of $\frac{.10 * \hat{x}}{\sigma_x} = \frac{1}{3}$, i.e., a $\frac{1}{3}$ standard deviation increase causes an increase of $10\%$ with respect to the mean of $x$. Secondly, assume this person also has the variables $\{w, y, z\}$ that affect $x$ (so $\vec{v} = \{w, x, y, z\}$). For now, we only consider one of these variables, namely $y$. Assume that we have calculated that a unit (one standard deviation) impulse in $y$ has a unit increase in $x$ over an arbitrary horizon of $10$ steps, i.e., $irf_{cum}(y,x,10) = 1$. Since the \textsc{irf} are standardized, this corresponds to a one standard deviation response. For each step, the effect of $y$ on $x$ is therefore on average $\frac{effect}{horizon}$. Recalling that the person would like an increase in $x$ of $10\%$, equal to a $\frac{1}{3}$ standard deviation increase on a single step. As $y$ has an average effect of $\frac{1}{10}$ standard deviation on $x$ per step, we require a difference of $\frac{3^{-1}}{10^{-1}} = \frac{10}{3}$ standard deviations in $y$. Because the impulse is standardized, we can determine the exact percentage of change needed in $y$ with respect to the average of $y$, $\hat{y}$. Thus, the required change was $\frac{10}{3} * \sigma_y$, which is $\frac{10 \cdot 3^{-1} * \sigma_y}{\hat{y}} * 100$ \% required change in $y$.


\section{Algorithms} 
\label{sec:implementation}
\textsc{Aira} is freely available (open source) from \url{http://frbl.eu/aira}. We implemented two versions of the proof-of-concept algorithm, one version in JavaScript (\url{https://github.com/frbl/aira}), a language designed to run client-side in a web browser, and a version in the R-language (\url{https://github.com/frbl/airaR}), a software environment for statistical computing. JavaScript has the advantage that the calculations can be performed on the client and do not require any computation on the server. JavaScript also provides interactivity in the client's browser; it allows for live updating the browser's \textsc{dom} (Document Object Model) of the web-page, enabling animations and interactivity. Furthermore, the JavaScript implementation could aid the use of \textsc{aira} on a large scale, as the implementation can be used on a back-end server (for example using NodeJS) or in the browser. A live demo can be found at \url{http://frbl.eu/aira}.

Each of the algorithms used for generating advice in \textsc{Aira} is elaborated in the following sections. The algorithms used for performing the \textsc{irf} calculations and for converting the \textsc{var} model to a \textsc{vma} representation are provided in \appref{app:impulse_response_calculation}. In all examples, $m$ denotes the number of variables in the model and $p$ the number of lags.

 \subsection{Selecting Variables and Determining Advice}
 \label{sub:determining-advice}
 The advice generation of \textsc{aira} is split up into three parts: \rom{1} \q{most influential node in network} (determining the net effect of a variable on well-being), \rom{2} \q{length of effect} (presenting the duration of a significant effect), and \rom{3} \q{percentage effect} (giving advice on how each of these variables can actually be changed). The pseudocode of the algorithm used for the first type of advice is provided in \algoref{alg:aira_optmal_effect_function}. The result of this function is an associative array ($r$) of cumulative effects (positive or negative) for each variable, sorted by descending absolute value. The algorithm iterates over all variables in the model (line $3$). From line $6$--$14$ it determines the net cumulative \textsc{irf} effect a variable has on each other variable, and stores summed effect per variable in $r$.
  
\begin{algorithm}[ht]
    \begin{algorithmic}[1]
        \Function{DetermineMostInfluentialNode}{$C, k$}
            \LineComment{$C$ are the VMA coefficients as returned by \algoref{alg:vma_coef}, $k$ is the horizon to forecast.}
            \State{$r\gets \text{associative array}$}
            \For{$ x\gets 1, x \le m$}
                \State{$r_x \gets 0$}
                \State{$eff \gets \Call{CalculateIRF}{\vec{\alpha}(x), C, k}$}
                \For{$y\gets 1, y \le m$}
                    \If{$x \ne y$}
                        \For{$l\gets 1, l \le k$}        
                            \State{$r_x \gets r_x  + (eff_y)_l$}
                            \State{$l \gets l + 1$}
                        \EndFor{}
                    \EndIf{}
                    \State{$y \gets y + 1$}
                \EndFor{}
                \State{$x \gets x + 1$}
            \EndFor{}
            \State{$r \gets \text{Sort } r \text{ based on the absolute values}$}
            \State{\textbf{return} $r$}
        \EndFunction{}
    \end{algorithmic}
    \caption{Determines the most influential variable.}
    \label{alg:aira_optmal_effect_function}
\end{algorithm}

Besides showing the net effect a variable has on well-being, \textsc{aira} shows how long an impulse has a (significant) response. \textsc{Aira} makes it insightful what the effect duration would be when an impulse is given to another variable. The algorithm iterates over all steps on the horizon (line $7$--$22$). For each step it then checks whether the effect differs (significantly) from zero (line $8$). It does so until an effect has been found. When this first effect has been found, a flag ($effect\_started$) is toggled (line $12$), and the start and direction of the effect are estimated. The direction is either positive or negative, and is determined in line $9$. If the effect does not start in the first step, the effect is linearly interpolated to the expected moment it passed a threshold (line $11$). This continues until the effect converges or exceeds the threshold. If this is the case, the algorithm linearly interpolates the point where it exceeded the threshold (line $16$). This result is stored, in terms of the total duration of the effect (line $18$) and the total time the effect took to converge to zero (line $19$). In case the chosen horizon was too small, and the effect did not have enough time to converge, the total effect and effective horizon are determined in lines $23$--$26$. The algorithm returns several values on line $27$. Firstly it returns the total time of the effect, that is, the length of the effect multiplied by the interval between measurements, yielding the total length of the effect in minutes. Secondly it returns the total length of the effect, that is, the number of steps where the (significant) effect is non-zero. Lastly it returns the total effective horizon. That is, the total horizon over which the effect is non-zero (i.e., the last step where there is a (significant) effect. It defaults to the horizon if the effect does not converge within the given horizon).

\begin{algorithm}[ht]
  \begin{algorithmic}[1]
  \Function{DetermineLengthOfEffect}{$x, y, inter, k$}
    \LineComment{$x$ is the variable that receives the shock, $y$ is the variable we measure the response on, $inter$ is the interval with which the data was sampled, and $k$ the horizon to forecast.}
    \State{$\vec{\gamma} \gets \Call{irf}{x,y,k}$}
    \State{$start,\ end \gets 0$}
    \State{$effect\_started \gets FALSE$}
    \State{$total,\ effective\_horizon,\ d \gets 0$}
    \State{$threshold \gets 1e^{-4}$}
    \For{$ i\gets 1, i \le k$}
        \If{$\mid \vec{\gamma}_i \mid\ > threshold$}
            \State{$d \gets ((\vec{\gamma}_i > threshold)$ ? $-1 : 1)$}
            \If{$i > 1 \cap \neg effect\_started$}
                \State{$start = i - \frac{(\vec{\gamma}_i + d \cdot threshold)}{(\vec{\gamma}_i - \vec{\gamma}_{i - 1})}$} 
                \State{$effect\_started \gets TRUE$}
            \EndIf{}
        \Else{}
            \If{$effect\_started$}
                \State{$end = i - 1 + (1 - \frac{(\vec{\gamma}_i + d \cdot threshold)}{(\vec{\gamma}_i - \vec{\gamma}_{i-1})})$}
                \State{$effect\_started \gets FALSE$}
                \State{$total \gets total + (end - start)$}
                \State{$effective\_horizon \gets end$}
            \EndIf{}
        \EndIf{}
    \EndFor{}
    \If{$effect\_started$}
        \State{$total \gets total + (k - start)$}
        \State{$effective\_horizon \gets k$}
    \EndIf{}
    \State{\textbf{return} $[total \cdot inter,\ total,\ effective\_horizon]$}
  \EndFunction{}
  \end{algorithmic}
  \caption{Determines the actual effect length.}
  \label{alg:length_of_effect}
\end{algorithm}

Lastly, \textsc{aira} shows what a participant can do to adjust said variables. \algoref{alg:specific_advice} shows the pseudocode designed for generating this advice. The \textsc{avg} and \textsc{sd} functions used in the algorithm give respectively the average and the standard deviation of the values of the answers, as measured during the diary study. The algorithm iterates over all variables in the model (line $3$), and for each variable the algorithm determines the length of the effect, and the net effect the current variable ($y$) has on the variable to be changed ($x$) (lines $5$--$6$). In lines $7$ to $12$, this net effect is converted to a percentage of the average value of the variable and stored as a result, according to \eqnref{eq:percentage-effect}. The algorithm allows to filter out effects lower than a threshold $\theta$ (expressed in terms of standard deviations, line $7$), as the percentage needed for a change with a low effect might become unrealistically large.

\begin{algorithm}[htbp]
  \begin{algorithmic}[1]
  \Function{DeterminePercentageEffect}{$perc, x, \theta, k$}
    \LineComment{$perc$ is the percentage with which the variable to change ($x$) needs to be changed. $\theta$ is a threshold of minimal effect needed in the variables, and $k$ the horizon to forecast.}
    \State{$results\gets \text{associative array}$}
    \For{$y \gets 1, y \le m$}
      \If{$y \ne x $}
        \State{$\hat{k} \gets \Call{DetermineLengthOfEffect}{y, x, 0, k}_3$}
        \State{$n\_e \gets \Call{irf$_{cum}$}{y, x, \hat{k}}$}
        \If{$n\_e > \theta$}
        \State{$\Delta \gets perc\cdot 100^{-1}$}
        \State{$\zeta \gets \Call{avg}{x} \cdot \hat{k} \cdot \Delta \cdot \Call{sd}{y}$}
        \State{$\zeta \gets \zeta \cdot (n\_e^{-1} \cdot \Call{avg}{y}^{-1} \cdot \Call{sd}{x}^{-1})$}
        \State{$results_y \gets \zeta \cdot 100 $}
        \EndIf{}
      \EndIf{}
      \State{$y \gets y + 1$}
    \EndFor{}
    \State{\textbf{return} $results$}
  \EndFunction{}
  \end{algorithmic}
  \caption{Determines the percentage of change needed per variable in order to induce a change of a certain percentage in another variable.}
  \label{alg:specific_advice}
\end{algorithm}


\subsection{Time complexity}
\label{sub:time_complexity}

We determined the time complexity of each of the algorithms presented in \secref{sec:implementation} using the \emph{Big-O} notation technique (denoted using $\mathcal{O}$). The time complexity of the algorithms for generating the \textsc{irf} functions are provided \appref{app:time_complexity}. These time complexities describe the upper bound of the processing time of the algorithm when the input size grows infinitely large.

For determining the most influential node in the model (\algoref{alg:aira_optmal_effect_function}), the algorithm calculates the total effect each variable has on all other variables and therefore loops over all $m$ variables in the model $m$ times. For each of these variables, it determines the \textsc{irf} once. Finally, it iterates over each calculated response on the horizon of length $k$. The total time complexity of determining the advice is therefore $\mathcal{O}(m \cdot (\mathcal{O}(\text{\textsc{CalculateIRF}}) + mk))$ or $\mathcal{O}(m \cdot (k^2m^2 + mk))$, which reduces to $\mathcal{O}(k^2m^3)$, where $m$ is the number of variables in the model and $k$ is the horizon of the \textsc{irf}.

For determining the length of the effect a variable $x$ has on a variable $y$ (\algoref{alg:length_of_effect}), the algorithm first calculates the \textsc{irf} of this effect ($\mathcal{O}(\text{\textsc{irf}})$). The algorithm itself then loops over each of the $k$ steps on the horizon ($\mathcal{O}(k)$). As calculating the \textsc{irf} has a higher complexity than $\mathcal{O}(k)$, the upper bound for the complexity of this algorithm is $\mathcal{O}(\text{\textsc{irf}}) = \mathcal{O}(k^2m^2)$.

Determining the percentage advice as listed in \algoref{alg:specific_advice} considers the effect all variables have on a single variable. Hence, for this algorithm we do not need to loop over all variables more than once, making the time complexity $\mathcal{O}(m \cdot (\mathcal{O}(\text{\textsc{DetermineLengthOfEffect}}) + \mathcal{O}(\text{\textsc{irf}}_{cum})) = \mathcal{O}(m \cdot (k^2m^2 + k^3m^2)$. When one would apply dynamic programming to cache the \textsc{irf} calculation, the time complexity could be reduced to $\mathcal{O}(2mk + k^2m^2) = \mathcal{O}(k^2m^2)$, as the \textsc{irf} then only has to be calculated once for all variables.

We can now define the total time complexity of \textsc{aira} as the maximum complexity of its algorithms. The algorithm with the highest complexity is the algorithm for calculating \algoref{alg:specific_advice} (\textsc{DeterminePercentageEffect}). This results in the time complexity of \textsc{aira} being $\mathcal{O}(m^{3}k^{3})$.

The algorithms presented in this section do allow for parallelization on several levels. For example, the most complex algorithm (\algoref{alg:specific_advice}) can be optimized by parallelizing its outer loop, in which it iterates over the $m$ variables in the model. This reduces the complexity of \textsc{aira} by a factor $m$, resulting in a time complexity of $\mathcal{O}(m^{2}k^{3})$ for \textsc{aira} on systems with at least $m$ threads available for parallel execution.

The practical performance of \textsc{aira} is acceptable and usable for general use. For instance, calculating advice for a model of six variables and a horizon of twenty steps took less than a second, as measured on both a modern laptop and a tablet.

\section{Experimental results}
\label{sec:experimental_results}
We performed several experiments to evaluate the effectiveness and accuracy of \textsc{aira}. These experiments show possible use cases of \textsc{aira} and give an impression of the three advice types \textsc{aira} can generate. We ran each of the aforementioned algorithms on data sets from two studies. Furthermore, we compared \textsc{aira}'s results of the \textsc{LengthOfEffect}-algorithm to the results of an earlier study.

The first dataset we used originated from the HowNutsAreTheDutch study, from which we randomly selected five participants having more than $85\%$ or $77$ completed measurements ($N=164$, henceforth referred to as the HND data set)~\cite{VanderKrieke2015}. For these five participants, we fitted \textsc{var} models using the AutovarCore procedure~\cite{Emerencia2016-AutovarCore}. AutovarCore is a faster and more efficient version of the original Autovar procedure~\cite{Emerencia2013, Emerencia2016-AutovarCore}. We used three variables recorded in this study: \rom{1} \emph{feeling gloomy}, \rom{2} \emph{relaxation}, and \rom{3} \emph{feeling inadequate}. These variables were selected so our experiment contains both positive and negative variables.

The second data set was retrieved from the study performed by \citeauthor{RefWorks:4} (henceforth referred to as the Rosmalen data set)~\cite{RefWorks:4}. \citeauthor{RefWorks:4} investigate the relation between depression and activity using \textsc{var} and \textsc{irf} analysis. In their work, they describe for one subject how long significant effects on activity and depression remain by inspecting the \textsc{irf}.
 
\subsection{Most influential node}
\label{sub:total_effect}
We ran \algoref{alg:aira_optmal_effect_function} on the HND data set to demonstrate a particular use case of \textsc{aira}. We loaded each model in \textsc{aira} and applied the procedure as described in \algoref{alg:aira_optmal_effect_function}, \textsc{DetermineMostInfluentialNode}. We marked \q{feeling gloomy} and \q{feeling inadequate} as negative variables and only used effects having a confidence of at least $95\%$ by bootstrapping the results $200$ times. The results are shown in \tabref{tab:effects_in_aira}. Note that the \q{feeling gloomy} and \q{feeling inadequate} variables have been converted to positive variables (resp. \q{feeling less gloomy} and \q{feeling less inadequate}).

\input{output/tab_effects_in_aira.tex}

These results show several interesting findings. First of all, they emphasize the difference between the participants. Some participants show similar responses, but some participants are rather deviant or even have opposite responses. For person $2$ and person $5$ \q{feeling less gloomy} seems to have a positive effect on \q{well-being}, whereas the effect is neutral for the other participants. For person $1$ \q{relaxation} seems to have a negative effect on \q{well-being} whereas this effect is positive for person $5$. \q{Feeling less inadequate} seems to have a small positive effect for person $1$ and person $3$. The variables do not seem to have an effect on the well-being of person $4$. 

\subsection{Length of the effect}
\label{sub:length_of_the_effect}
In the second evaluation, we compared \textsc{aira}'s method for determining the length of an effect to the results of the manual analysis by \citeauthor{RefWorks:4}~\cite{RefWorks:4} (the Rosmalen data set). In this experiment, \textsc{aira} was configured to use orthogonalized \textsc{irf} similar to \citeauthor{RefWorks:4} We compared the contemporary effects where activity is assumed to precede depression with those of \citeauthor{RefWorks:4} (see~\cite{RefWorks:4} for more information).

During the experiment, we noticed differences between our outcomes and those of \citeauthor{RefWorks:4} These differences mainly originate from differences in the statistical packages used. Firstly, there is a small difference between the \textsc{var} coefficients generated for \textsc{aira} (in R using the \emph{vars} package~\cite{Pfaff2008}) and the models used by \citeauthor{RefWorks:4} (as generated using Stata 11.0~\cite{Stata11}). Secondly, there seem to be differences in the confidence intervals for \textsc{irf} as calculated in Stata compared to those calculated using the R \textsc{vars} package. These differences seem to be caused by the different methods used by Stata and R to calculate the \textsc{var} models and error bands for \textsc{irf} (see~\cite{Pfaff2008} and the Stata Time series Manual for details). Lastly, the method for calculating the confidence intervals uses bootstrapping, which depends on a random component that also contributes to the differences.

To remove the bias of differing statistical packages, we refitted each of the models from \citeauthor{RefWorks:4} using the R \textsc{vars} package and compared \textsc{aira} with these models. The \textsc{irf} time plots of these models are shown in \appref{app:time_plots} and are nearly identical to the results of \citeauthor{RefWorks:4}, as shown in Figure 2 on page 7 of~\cite{RefWorks:4}.

\tabref{tab:comparison} shows the outcome of \textsc{aira}. The first two columns (annotated with (1)) show the predictions of \textsc{aira}, the third and fourth column (annotated with (2)) show the predictions based on the \textsc{var} models in the work of \citeauthor{RefWorks:4}, and the last two columns (annotated with (3)) show the results of a manual inspection of the \textsc{var} models from \citeauthor{RefWorks:4} refitted using the \textsc{vars} package. The results between \textsc{aira} and the refitted models are very similar, and it can be argued that the results of \textsc{aira} are in fact more precise as \textsc{aira} applies linear interpolation to estimate the exact length of the effect.

\input{output/tab_comparison.tex}


\subsection{Percentage effect}
\label{sub:percentage_effect}
In the third evaluation, we evaluate \algoref{alg:specific_advice} using the HND data set. For each of the modeled variables, we determined how well each of the other variables in the model could be used to increase the positive variables or decrease the negative variables by $10\%$. That is to say, we determined how well \q{activity} and \q{relaxation} can be used to reduce \q{feeling nervous} by $10\%$. If the found effect was negligible, or if there was no effect at all, the variables were assumed to require an infinite increase in order to change the variable. For each of the results we only used the effects having a certainty of at least $95\%$ (by bootstrapping $200$ times). The results of the algorithm are (only showing percentages $\ge -1000\%$ and $\le 1000\%$): \input{output/item_percentage_effects_in_aira} In this example, \textsc{aira} was able to generate advice for person $2$ and person $5$. These results show that for person $2$, in order to decrease \q{feeling inadequate} by $10\%$, he or she would need to feel approximately $90\%$ less gloomy than now. Person $5$ could decrease his or her feeling of \q{inadequacy} by $10\%$, by \q{relaxing} approximately twice as much. Although these outcomes might be hard to interpret in the present form, they can give an idea which of the variables would be the best fit for intervening on another variable.

\section{Discussion} 
\label{sec:discussion}
\textsc{Aira} provides tailored advice about mental health and well-being, using data collected in a diary study. \textsc{Aira} allows participants to inspect how relations between their psychological, and how they interact over time by means of \textsc{irf}. We performed several experiments to demonstrate the performance of \textsc{aira}. In our experiment, presented in \secref{sub:length_of_the_effect}, we showed that the results from the automated analysis of \textsc{aira} are comparable to earlier work and arguably more precise.

In mental health research, few case studies (generally consisting of one to five participants) have previously applied \textsc{irf} analysis to diary data  (e.g., \citeauthor{RefWorks:4}~\cite{RefWorks:4} and \citeauthor{Hoenders2011}~\cite{Hoenders2011}). \textsc{Aira} includes several methods to perform an \textsc{irf} analysis similar to the one used previously, but instead of using manual analysis, \textsc{aria} applies an automated technique. To the best of our knowledge, \textsc{aira} is the first approach for automatically generating \textsc{irf}-based advice.

The analysis of \textsc{irf} and \textsc{var} models leaves room for ample discussion. Firstly, a point of discussion is whether or not to include contemporaneous effects in \textsc{irf} analysis~\cite{RefWorks:5, Brandt2007}. Because the directionality of contemporaneous effects has no foundation in reality and therefore lacks conceptual justification, the use of contemporaneous effects is controversial. Although some of these directions can be determined using theoretical domain knowledge, this might not hold for the individual. Therefore, \textsc{aira} currently does not base its advice on contemporaneous effects.

Secondly, the use of \textsc{irf} and \textsc{var} models in the context of psychopathology research can be a point of discussion. Both \textsc{var} models and \textsc{irf} analysis are not originally developed to be used with \textsc{ema} data. In this study, we assume that a \textsc{var} model fit to \textsc{ema} data is representative for a period that exceeds the \textsc{ema} study itself, and is representative for future events. Treating the model as a linear and time invariant system enables us to perform such analysis. We based this assumption on several previous studies also applying \textsc{var} analysis to psychological research data (e.g., ~\cite{RefWorks:4, VanGils2014, Hoenders2011}). We should note that, as for many statistical techniques, a key requirement for \textsc{aira} to work is that the model comprises relevant variables. When the model merely consists of variables not having meaningful connections, not having meaningful variance, or variables that cannot be influenced \textsc{aira} will not function properly.

Thirdly, when a \textsc{var} model has more than one lag, it might happen that the sign of a \textsc{var} coefficient (i.e., the direction of the effect) is not equal for all lags. For example, a variable may have a negative effect on another variable in one lag but a positive effect in a different lag. In this scenario, it is difficult to determine whether the effect of the variable should be considered positive or negative when basing this decision merely on the \textsc{var} model. In \textsc{aira}, we circumvent this issue by using \textsc{irf}, and moreover, by summing the cumulative \textsc{irf} values to obtain a net effect value. This net effect value considers the entire horizon for determining an effect to be either a net gain or a net loss. That is, if the positive effects outweigh the negative effects, the relation is considered positive (and vice versa for negative effects). This approach is novel and specific to \textsc{aira}.

Fourthly, the creation of the \textsc{var} model. Although the creating of \textsc{var} models is deliberately left out of the present work, there are a few important points one needs to take into account. The estimation of a \textsc{var} model entails various caveats, such as the selection of the correct lag length and the equidistant measuring of the \textsc{ema} data set. When fitting a \textsc{var} model to data, these are properties to take into account. For the present work, we relied on previous work allowing the creation of \textsc{var} models to be performed automatically~\cite{Emerencia2013}.

The performance of \textsc{aira} with respect to the calculation time is sufficient for practical use. Basic experiments show that the calculation of an advice using a \textsc{var} model consisting of six variables and a horizon of twenty steps happens in less than a second on a modern laptop. Furthermore, our time complexity analysis implies that this performance scales well within acceptable boundaries for practical use. \textsc{Aira} is implemented in the R-language and in JavaScript, and can therefore run in any browser on any modern operating system (mobile or desktop). It should be noted that the running time for computing advice using bootstrapped \textsc{irf} analysis increases linearly with the (constant) number of bootstrap iterations.

Objectively establishing the correctness of any algorithm that provides suggestions or advice is a complex issue. \textsc{Aira} forms no exception. While the implemented formulas may be mathematically correct, we have not yet evaluated the practical utility of the advice generated by \textsc{aira} in terms of clinical accuracy, understandability, and helpfulness for the individual. Hence, whether the advice actually contributes to the well-being of an individual remains to be shown in future research.

In a broader perspective, \textsc{aira} could be a useful asset in the current field of mental health research and clinical practice. \textsc{Aira} could be used on large scale platforms as a decision support tool, and as a means to give automatic personalized advice on mental health and well-being, perhaps on a national scale such as in the HowNutsAreTheDutch project~\cite{VanderKrieke2015}.


\section{Conclusion and Future Work} 
\label{sec:conclusion_and_future_work}
The present work describes the Automatic Impulse Response Analysis (\textsc{aira}), an algorithm and related implementations on two platforms that can automatically provide feedback and advice on time series data such as diary data collected using the ecological momentary assessment method. Our paper described the basis and theoretical foundation of \textsc{aira}. We created a method to give specific advice on which variables require change, and with what percentage, in order to have the desired adjustment in other variables. Furthermore, we provided a proof-of-concept implementation of this algorithm for use in e-mental health platforms. \textsc{Aira} provides an automated method to deal with the previously unsolved problem where the lags of a variable have conflicting effects (e.g., a positive effect on one lag and a negative effect on another lag). These mixed effects make it difficult to determine whether the overall effect of a variable is a net gain or a net loss. Using \textsc{aira}, these effects can be summarized, revealing the net effect.

Future versions of \textsc{aira} could include algorithms to determine which sequence of impulses over time, rather than which single impulse, has the desired effect. Moreover, the percentages currently provided could be converted to an easier to understand format, such as time investment required to effectuate a change. In its current form, \textsc{aira} can be used by participants to find out how best to self-manage their well-being. This can be improved by allowing the individual to personally assign importance to the variables under study. We currently apply a general, binary approach to determining whether a variable is considered positive or negative. This self-management can be improved further by allowing users to provide a relative importance for each variable.



\section*{Acknowledgments}
\label{sec:acknowledgments}
  \textsc{Aira} is developed for the Leefplezier and the How\-Nuts\-Are\-The\-Dutch project. The Leefplezier project is funded by Espria. Espria is a healthcare group in the Netherlands consisting of multiple companies targeted mainly at the elderly population. The How\-Nuts\-Are\-The\-Dutch project is partly funded by NWO VICI grant no.~91812607, awarded to Prof.\ Peter de Jonge. We thank all people working on How\-Nuts\-Are\-The\-Dutch, Leefplezier, and RoQua. Especially, the authors thank Elske Bos, for providing invaluable input in the design of \textsc{aira}.

\bibliographystyle{IEEEtranN}
\bibliography{library}

\appendix

\section{Impulse response calculation} 
\label{app:impulse_response_calculation}
\textsc{Aira} first converts \textsc{var} models into the \textsc{vma} representation of the model. \algoref{alg:vma_coef} shows the pseudocode for determining the \textsc{vma} coefficients. In this (and the other) examples, $m$ denotes the number of variables in the model and $p$ the number of lags. The R version of \textsc{aira} uses the \emph{vars} package to calculate these models~\cite{Pfaff2008}. The JavaScript version of \textsc{aira} uses an implementation specially crafted for the present work. This JavaScript implementation for calculating the \textsc{vma} models and running the \textsc{irf} analysis is described in this appendix.

Firstly, the \textsc{var} coefficient matrix (of size $m \times mp$) in the \textsc{var} model is partitioned in separate matrices based on the lag, generating $p$ matrices of size $m\times m$, as shown in lines $4$--$8$ of \algoref{alg:vma_coef}. This allows for reusing them when determining the \textsc{vma} coefficients. Secondly, \textsc{aira} performs the transformation from \textsc{var} coefficients (the $B$ matrices in the \textsc{var} equation) to the \textsc{vma} coefficients (the $C$ coefficient matrix in \textsc{vma} representation), as shown in lines $9$--$17$. Note that only the first $k$ rows of $C$ are converted, i.e., the specified horizon.

The $\delta$ function used in this algorithm in line $9$, $12$, and $15$ checks whether a coefficient matrix is available for the provided lag (i.e., a model with two lags has two coefficient matrices, one for each lag). This function is described in \eqnref{eq:delta_func}. The algorithm for the \textsc{vma} coefficients is based on the work of \citeauthor{Brandt2007}~\cite{Brandt2007} and on the work of \citeauthor{RefWorks:3}~\cite{RefWorks:3}. 

\begin{algorithm}[ht]
  \begin{algorithmic}[1]
    \Function{CalculateVMA}{$var\_coef, p, k$}
        \LineComment{$var\_coef$ is the coefficient matrix from the VAR model (of size $m\times mp$), $p$ is the number of lags, and $k$ is the number of steps to forecast (the horizon).}
        \State{$B\gets \text{empty list of size }p$}
        \State{$C\gets \text{matrix of all zeroes of size }(k \times k)$} \Comment{the \textsc{vma} coefficient matrix}
        \For{$l\gets 1, l \le p$}
            \State{$x \gets m \cdot (l-1) + 1$}
            \State{$B_l \gets var\_coef_{(1 \dots m), (x \dots x+m-1)} $}\Comment{$B$ is a list of matrices, each matrix being the coefficients for a different lag}
            \State{$l \gets l + 1$}
        \EndFor{}{}
        \State{$C_{1,1} \gets \Call{$\delta$}{B, 1}$}
        \For{$ i\gets 2, i \leq k $}
            \For{$j\gets 1, j < i$}
                \State{$C_{i,j} \gets \delta(B, j) \cdot (\sum_{x=1}^{i-j} C_{i-j,x})$}
                \State{$j \gets j + 1$}
            \EndFor{}{}
            \State{$C_{i,i} \gets \Call{$\delta$}{B, i}$}
            \State{$i \gets i + 1$}
        \EndFor{}{}
        \State{\textbf{return} $C$}
    \EndFunction{}
  \end{algorithmic}
  \caption{Finds the VMA coefficients of the VAR model.}
  \label{alg:vma_coef}
\end{algorithm}

\begin{equation}
  \delta(B, j) =
  \begin{cases}
    m \times m\ \text{zero matrix}   & \text{if } j \ge B.length\\
    B_j                              & \text{otherwise}
  \end{cases}
  \label{eq:delta_func}
\end{equation}

After \textsc{aira} has calculated the vector moving average model, it performs the \textsc{irf} analysis. \algoref{alg:irf_func} lists the pseudocode used for running this analysis. The algorithm starts by defining an output matrix in line $2$. In the case of orthogonalized \textsc{irf}, the identity matrix used in line $3$ ($I_m$) should be replaced by the contemporaneous coefficient matrix. Orthogonalized \textsc{irf} is currently only supported in the R version of \textsc{aira}. The actual impulse responses are defined in line $4$--$11$. The shocks are multiplied with each of the \textsc{vma} coefficients to determine the response and are calculated for each moment on the horizon.

\begin{algorithm}[ht]
  \begin{algorithmic}[1]
    \Function{CalculateIRF}{$\vec{s}, C, k$}
      \LineComment{$\vec{s}$ vector of length $m$ containing the shocks, $C$ are the VMA coefficients as returned by \algoref{alg:vma_coef}, and $k$ is the number of steps (horizon) to forecast.}
      \State{$Y\gets \text{empty matrix of size }(k \times m)$}
      \State{$Y_1 = I_{m} \cdot \vec{s}$}
      \For{$ t\gets 2, t \leq k$} 
        \State{$Y_t \gets [0_{1}, \dots, 0_{m}]^T$}
        \For{$i\gets 1, i < t$}
          \State{$Y_t \gets Y_t + (C_{t - 1, i} \cdot \vec{s})$}
          \State{$i \gets i + 1$}
        \EndFor{}
        \State{$t \gets t + 1$}
      \EndFor{}
      \State{\textbf{return} $Y^T$}
    \EndFunction{}
  \end{algorithmic}
  \caption{Algorithm for calculating the \textsc{irf} from shocks in $\vec{s}$, for $k$ steps in the future.}
  \label{alg:irf_func}
\end{algorithm}

\begin{figure*}[htbp!]
\resizebox{\textwidth}{!}{
    \centering
    \subfloat[PP1 Activity shocked]{\label{fig:pp1-a}\includegraphics[width=.25\textwidth]{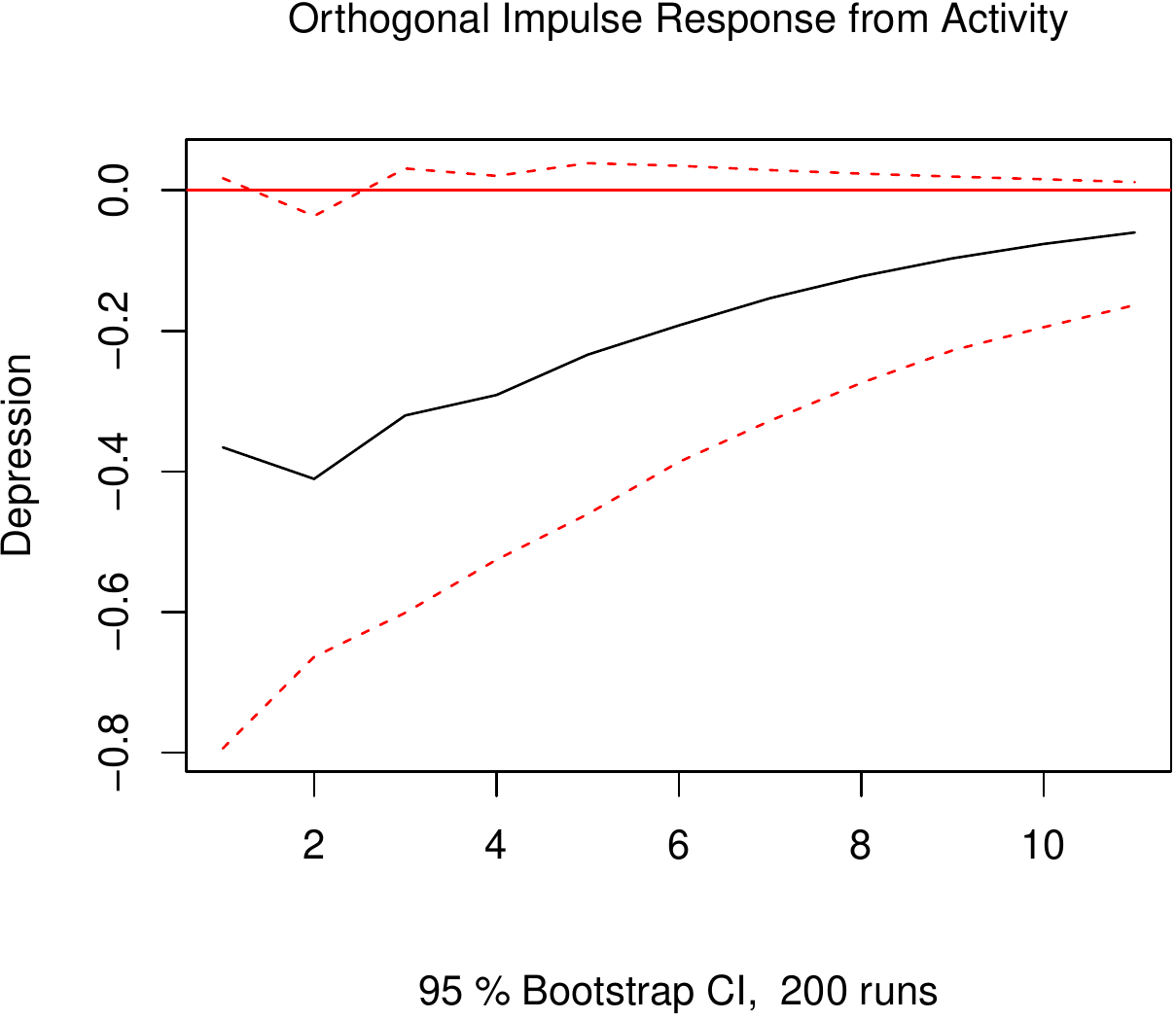}%
    }%
    ~
    \subfloat[PP2 Activity shocked]{%
        \includegraphics[width=.25\textwidth]{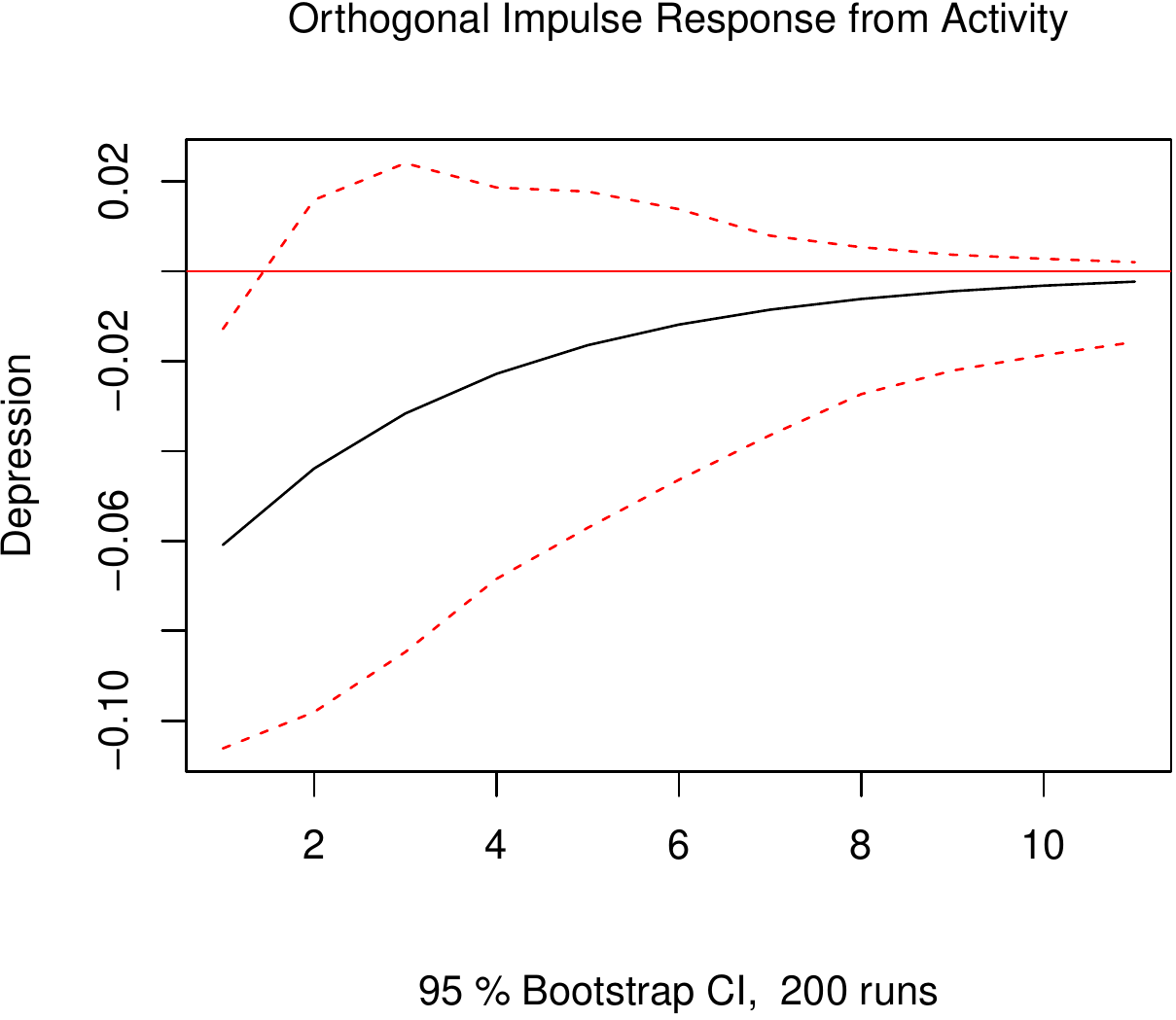}%
        \label{fig:pp2-c}%
    }%
    ~
    \subfloat[PP4 Activity shocked]{%
        \includegraphics[width=.25\textwidth]{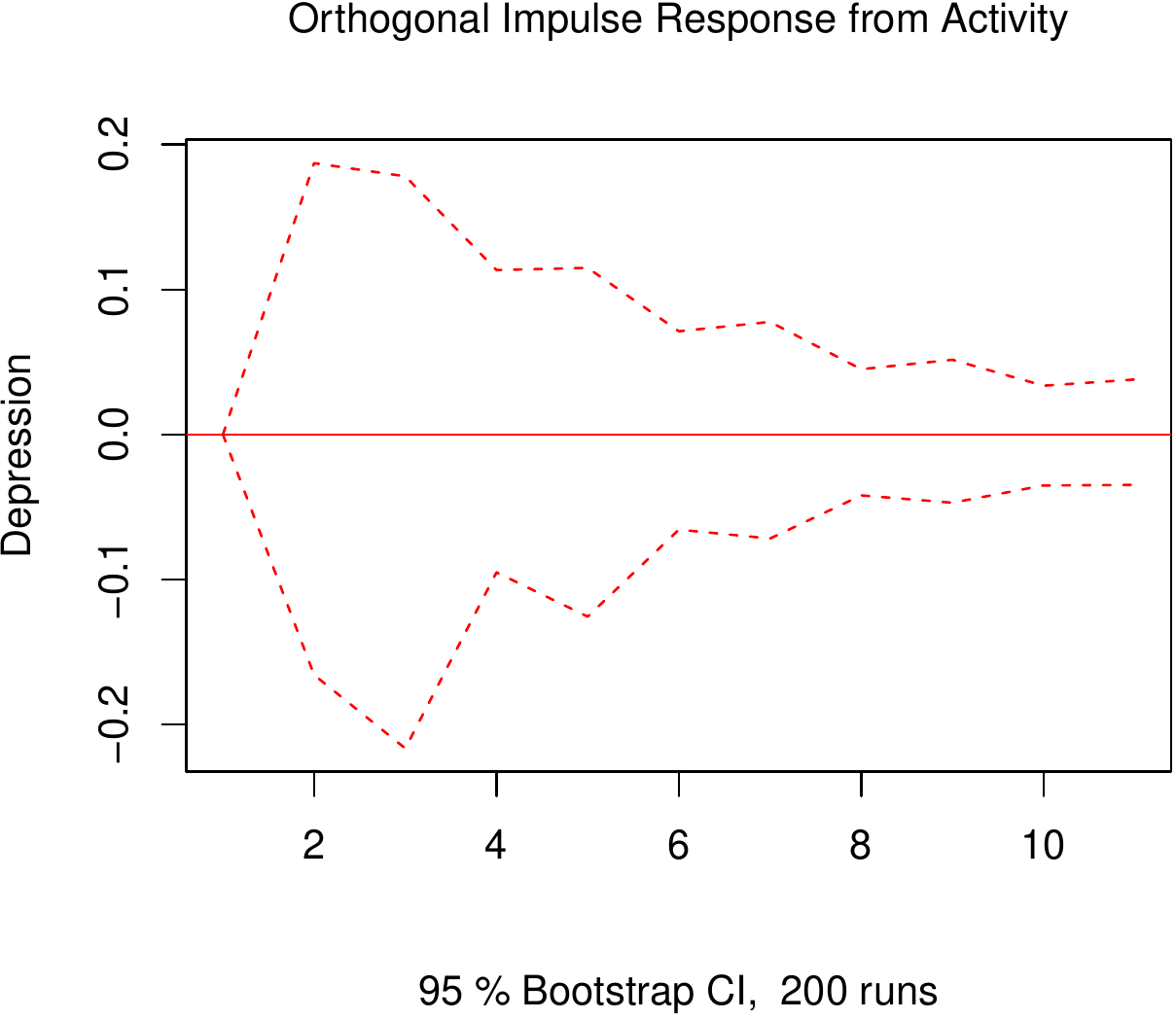}%
        \label{fig:pp4-a}%
    }%
    ~
    \subfloat[PP5 Activity shocked]{%
        \includegraphics[width=.25\textwidth]{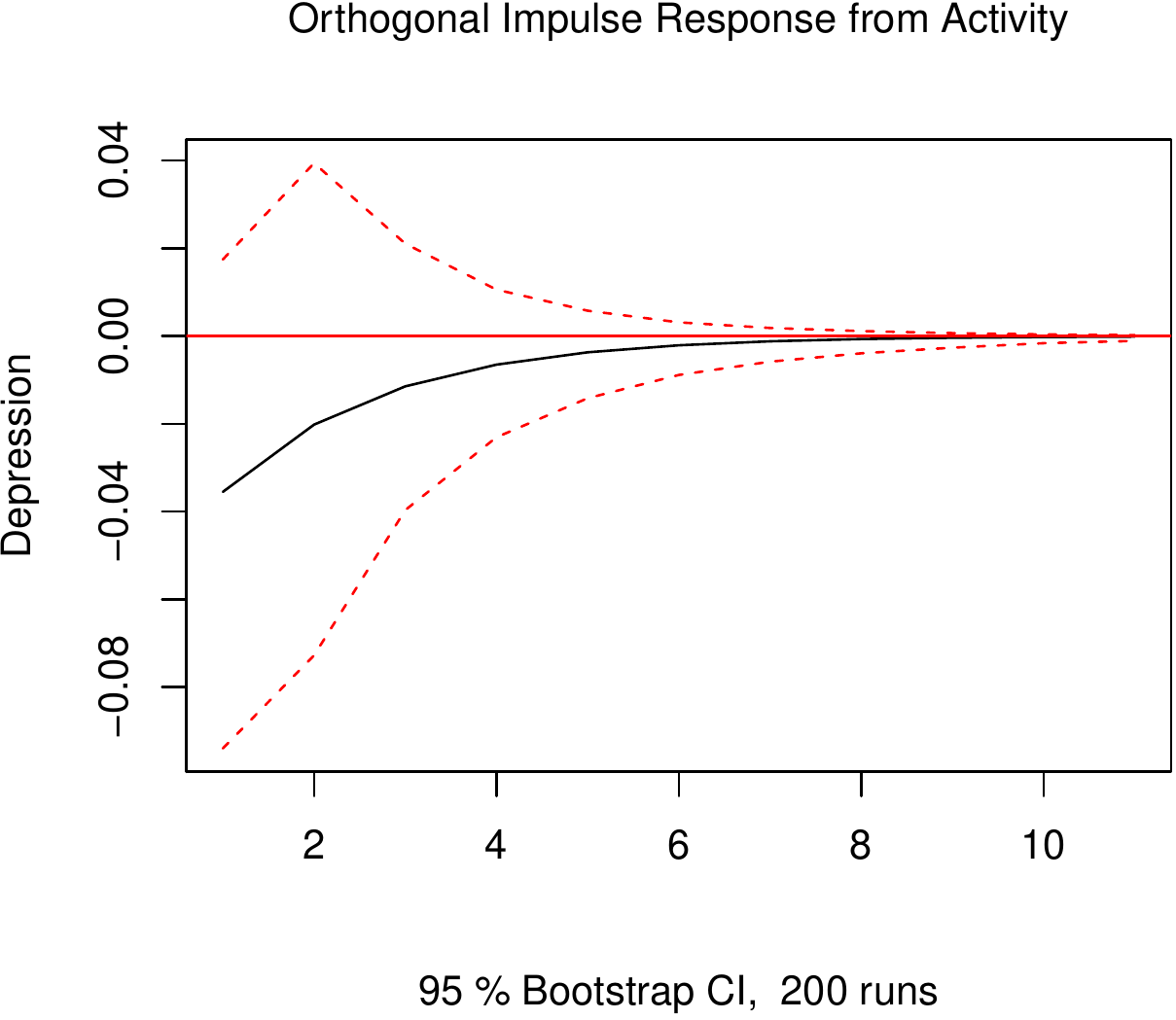}%
        \label{fig:pp5-c}%
    }%
    }
    \qquad
    \resizebox{\textwidth}{!}{
    \subfloat[PP1 Depression shocked]{\label{fig:pp1-b}\includegraphics[width=.25\textwidth]{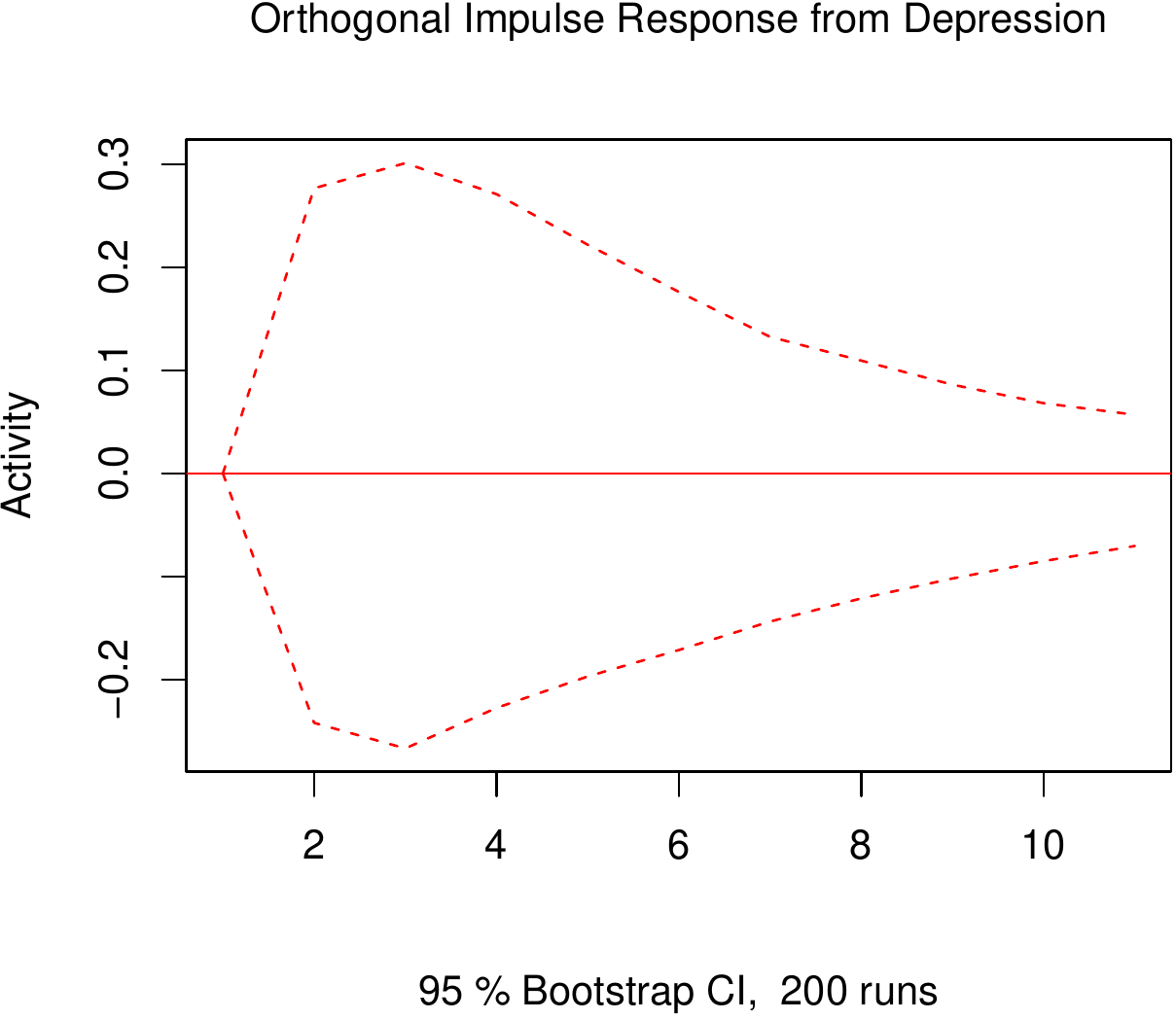}%
    }%
    ~
    \subfloat[PP2 Depression shocked]{%
        \includegraphics[width=.25\textwidth]{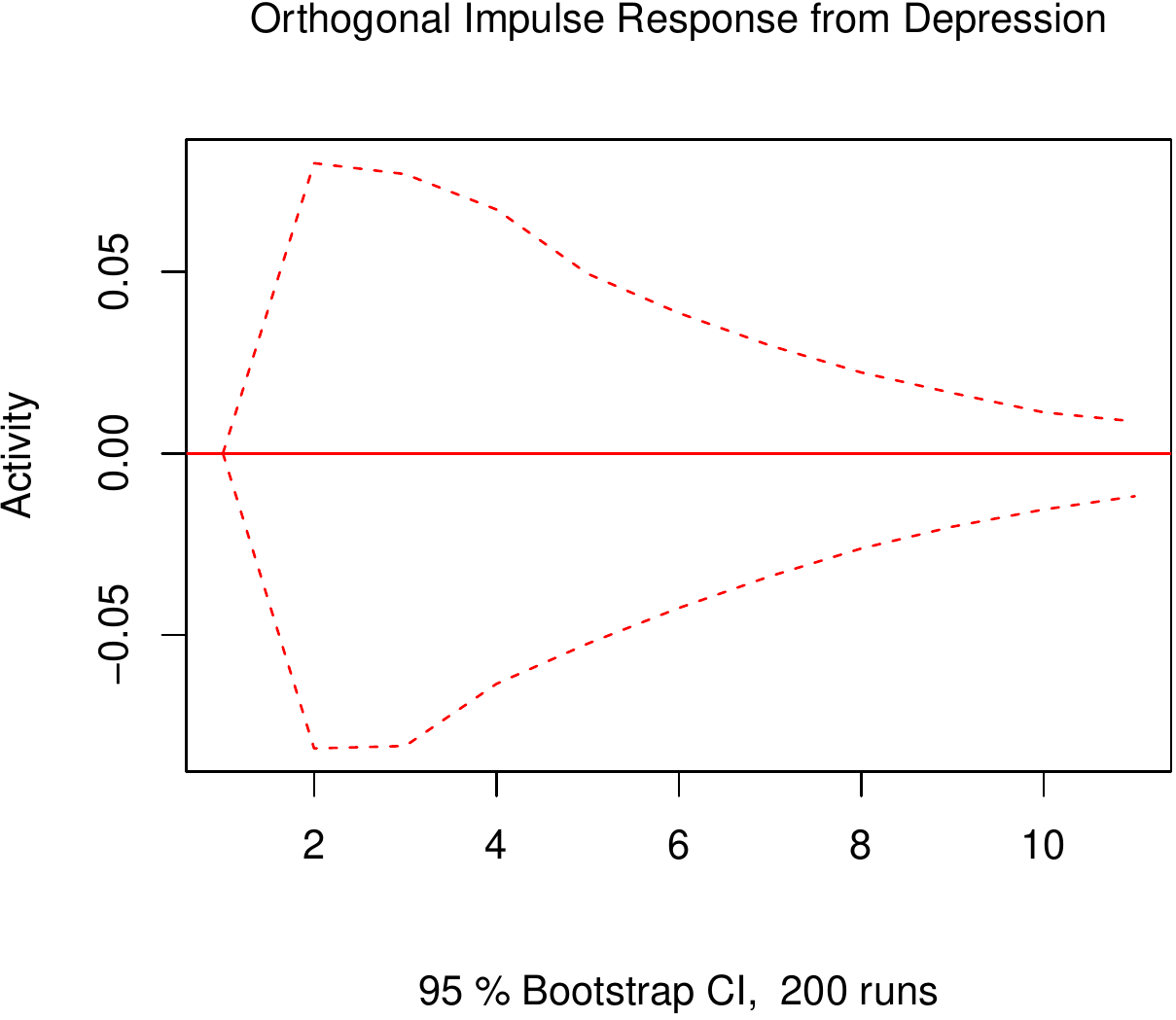}%
        \label{fig:pp2-d}%
    }
    ~
    \subfloat[PP4 Depression shocked]{%
        \includegraphics[width=.25\textwidth]{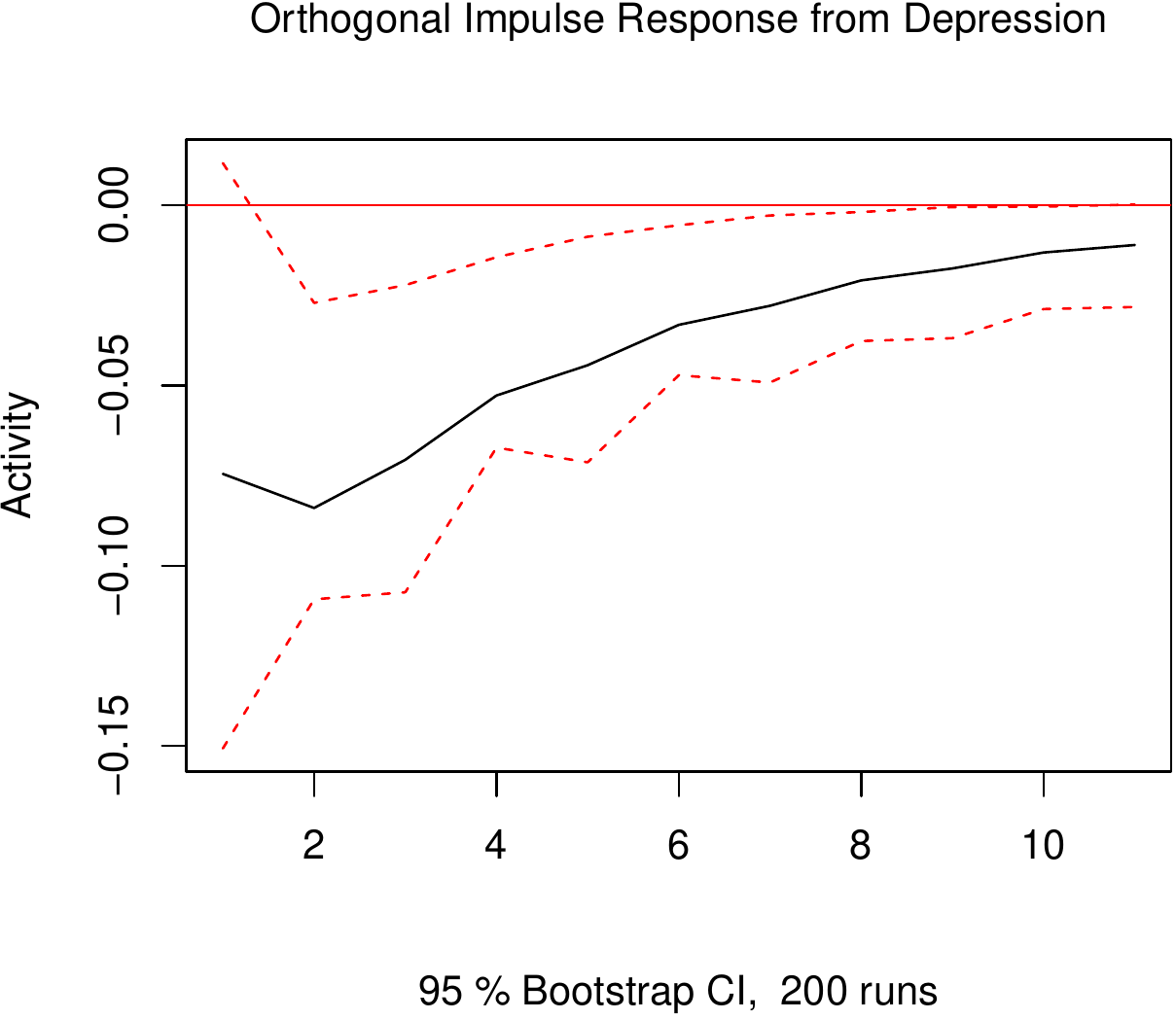}%
        \label{fig:pp4-b}%
    }%
    ~
    \subfloat[PP5 Depression shocked]{%
        \includegraphics[width=.25\textwidth]{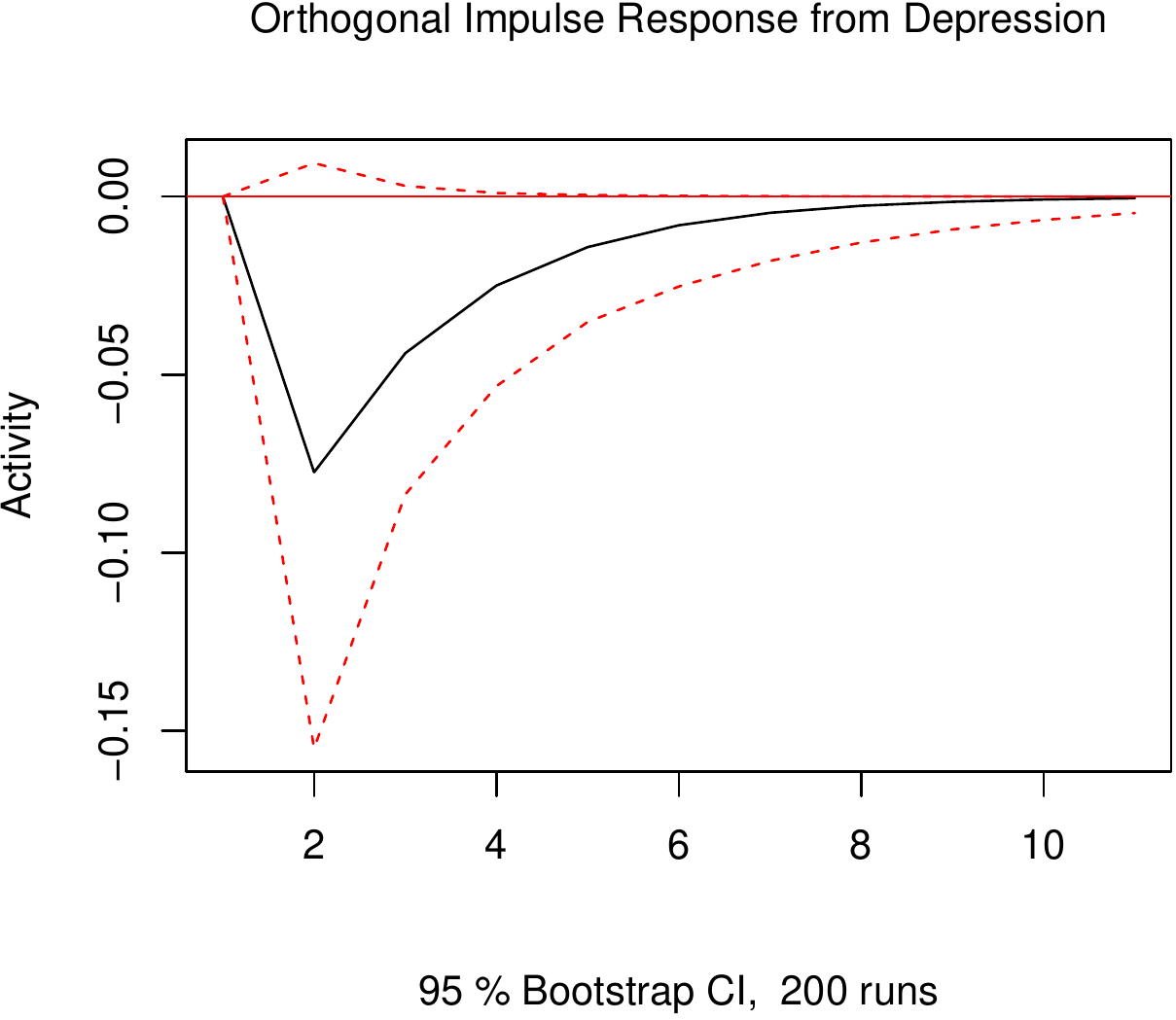}%
        \label{fig:pp5-d}%
    }
    }
  \caption{The \textsc{irf} output of the models from~\cite{RefWorks:4}, recalculated using the R vars package~\cite{Pfaff2008}. The red/dashed lines indicate the 95\% confidence interval, the black line the \textsc{irf} curve. Each of the examples has been bootstrapped $200$ times. The area under the 95\% confidence curves correspond to values presented in the main article.}
\label{fig:case-study-results}
\end{figure*}

\section{Time complexity}
\label{app:time_complexity}
\algoref{alg:vma_coef} describes the conversion of \textsc{var} coefficients to \textsc{vma} coefficients. In order to determine these \textsc{vma} coefficients, the algorithm iterates over $k$ (the used horizon). In each iteration, the algorithm retrieves all previously created \textsc{vma} coefficients $k$ times. This is done for all entries and therefore bounded by $k$. Each of the entries retrieved requires a matrix summation for each $m\times m$ matrix. Finally, this summed matrix is multiplied by another $m\times m$ matrix. The total time-complexity therefore has an upper bound of $\mathcal{O}(k^2m^3 + k^3m^2)$.

The time complexity of the \textsc{irf} calculation as shown in \algoref{alg:irf_func} depends on the horizon and the number of variables in the model. The algorithm iterates over the horizon ($k$ steps). For each step on the horizon, it determines an effect at most $k$ times, where each effect calculation is a matrix-vector multiplication of at most $m\times m$ steps (the size of matrix $C_{t-1,i}$). Finally, each element of the resulting $m \times 1$ vector is added to the $Y_t$ vector. The total upper bound of the calculation of the \textsc{irf} is therefore $\mathcal{O}(k^2m^2)$.

\section{Time plots}
\label{app:time_plots}
The plots of the impulse response functions used as input for the comparison of \textsc{aira} with the work of \citeauthor{RefWorks:4} are shown in \figref{fig:case-study-results}. In \figref{fig:case-study-results}, each \textsc{pp} represents a participant, the horizontal axis depicts the horizon of the \textsc{irf}, and the vertical axis shows the variable from which the response is recorded. \figref{fig:case-study-results} shows the \textsc{irf} graphs of four participants from \citeauthor{RefWorks:4}. These \textsc{irf} plots are similar to the time plots provided in the work of \citeauthor{RefWorks:4} (see Figure 2 on page 7 of~\cite{RefWorks:4}).

\end{document}

%% file: output/tab_effects_in_aira.tex
\begin{table}[ht]
\centering
\centering
\begin{tabular}{rrrr}
  \toprule
 & Feeling less gloomy & Relaxation & Feeling less inadequate \\ 
  \midrule
Person 1 & 0.000 & -0.061 & 0.045 \\ 
  Person 2 & 0.230 & 0.000 & 0.000 \\ 
  Person 3 & 0.000 & 0.000 & 0.057 \\ 
  Person 4 & 0.000 & 0.000 & 0.000 \\ 
  Person 5 & 0.015 & 0.410 & 0.000 \\ 
   \bottomrule
\end{tabular}
\caption{Effects of feeling less gloomy, relaxation and feeling less inadequate on well-being, in terms of standard deviations.} 
\label{tab:effects_in_aira}
\end{table}

%% file: output/tab_comparison.tex
\begin{table*}[ht]
\centering
\resizebox{\columnwidth}{!}{
\begin{tabular}{rllllll}
  \toprule
 & A $\rightarrow$ D (1) & D $\rightarrow$ A (1) & A $\rightarrow$ D (2) & D $\rightarrow$ A (2) & A $\rightarrow$ D (3) & D $\rightarrow$ A (3) \\ 
  \midrule
pp1 & 1746.2 minutes (1.2 days) & 0 minutes (0 days) & 5 days & 0 days & 1 days & 0 days \\ 
  pp2 & 639.4 minutes (0.4 days) & 0 minutes (0 days) & 3 days & 0 days & 1 days & 0 days \\ 
  pp4 & 0 minutes (0 days) & 11081.7 minutes (7.7 days) & 2 days & 6 days & 0 days & 7 days \\ 
  pp5 & 0 minutes (0 days) & 0 minutes (0 days) & 0 days & 1 days & 0 days & 0 days \\ 
   \bottomrule
\end{tabular}
}
\caption{Comparison between the outcomes of AIRA (1) and results from the study by \citeauthor{RefWorks:4} (2,3)~\cite{RefWorks:4}. The table shows both the results from the paper (2) and from the Rosmalen VAR models fitted using the VARS package (3).} 
\label{tab:comparison}
\end{table*}

%% file: output/item_percentage_effects_in_aira.tex
\emph{Person 2 can decrease feeling inadequate by changing feeling gloomy by -89.89\%, Person 5 can decrease feeling inadequate by changing relaxation by 206.38\%.}